\documentclass[prd,twocolumn,aps,amsmath,amssymb,superscriptaddress]{revtex4-1}

\usepackage{graphicx}
\usepackage{dcolumn}
\usepackage{bm}
\usepackage{wrapfig}
\usepackage{amsmath,amsfonts,amssymb}
\usepackage[usenames]{color}
\usepackage{bm}
\usepackage{slashed}
\usepackage{mathrsfs}
\usepackage{mathtools}
\usepackage{graphicx}
\usepackage{dcolumn}
\usepackage{bm}
\usepackage{bbm}

\newcommand{\be}{\begin{equation}}
\newcommand{\ee}{\end{equation}}

\usepackage{tikz}
\usetikzlibrary{patterns}
\usetikzlibrary{patterns.meta}
\usetikzlibrary{shapes,snakes}
\usetikzlibrary{calc}
\tikzset{font=\fontfamily{phv}\selectfont}
\usetikzlibrary{decorations.markings}
\tikzset{middlearrow/.style={
        decoration={markings,
            mark= at position 0.45 with {\arrow[black,
            scale=.5]{stealth}} ,
        },
        postaction={decorate}
    }
}
\usetikzlibrary{decorations.pathmorphing}
\tikzset{snake it/.style={decorate, decoration=snake}}

\begin{document}

\definecolor{purp_bg}{RGB}{210, 191, 255}
\definecolor{purp_line}{RGB}{94, 25, 255}
\definecolor{o1}{RGB}{250, 165, 7}
\definecolor{o2}{RGB}{244, 141, 6}
\definecolor{o3}{RGB}{232, 91, 4}
\usetikzlibrary{fadings}
\tikzfading[name=fadr,top color=o1, bottom color=o3]

\title{Decoherence of a 2-Path System by Infrared Photons}

\author{Colby DeLisle}
\email[]{delislecolby@gmail.com}
\affiliation{Department of Physics and Astronomy, University of
British Columbia, 6224 Agricultural Rd., Vancouver, B.C., Canada V6T
1Z1}
\affiliation{Pacific Institute of Theoretical Physics, University of
British Columbia, 6224 Agricultural Rd., Vancouver, B.C., Canada V6T
1Z1}

\author{P.C.E. Stamp}
\email[]{stamp@phas.ubc.ca, stamp@tapir.caltech.edu}
\affiliation{Department of Physics and Astronomy, University of
British Columbia, 6224 Agricultural Rd., Vancouver, B.C., Canada V6T
1Z1}
\affiliation{Pacific Institute of Theoretical Physics, University of
British Columbia, 6224 Agricultural Rd., Vancouver, B.C., Canada V6T
1Z1}
\affiliation{Theoretical Astrophysics, Cahill, California Institute
of Technology, 1200 E. California Boulevard, MC 350-17, Pasadena CA
91125, USA}

\date{\today}

\begin{abstract}

We calculate the decoherence caused by photon emission for a
charged particle travelling through an interferometer; the
decoherence rate gives a quantitative measure of how much
``which-path" quantum information is gained by the electromagnetic
field. We isolate the quantum information content of both leading
and sub-leading soft photons, and show that it can be extracted
entirely from information about the endpoints of the particle's
paths. When infrared dressing is used to cure the infrared
divergences in the theory, the leading order soft photons then give
no contribution to decoherence, and carry no quantum information.
The sub-leading soft photons in contrast may carry finite
which-path information, and the sub-leading contribution to
decoherence takes an extremely simple, time-independent form
depending only on the size of the interferometer.
An interesting open question is whether or
not dressing should also be applied at sub-leading order;
we discuss the possibility of answering this question experimentally.

\end{abstract}

\maketitle


\section{Introduction} \label{sec:intro}


Our subject in this paper is the well-known ``2-path" problem in
quantum mechanics \cite{feynman3}, in the present case for a charged
particle coupled to the electromagnetic (EM) field.  We thus assume a
pointlike particle with electric charge $e$, which is travelling
through an interferometer, and is put into a superposition of
trajectories falling into two distinct classes \cite{2class}, which
we label $L$ and $R$. We will sometimes refer to this particle as an
electron, although it could of course be some other charged particle
of larger mass.

An external observer then has the choice of either (i) measuring
which path the particle has followed, or (ii) letting it simply pass
through the interferometer unmolested by any measuring system. As is
well known, in the former case no interference is seen between the 2
paths, whereas in the latter case, a standard interference pattern is
seen on any screen which intercepts the 2 paths.

However, it is also well known that on its way through the
interferometer, the particle couples to the electromagnetic (EM)
field, which may be either at temperature $T=0$ (in its ground state)
or at finite $T$ in the rest frame of the interferometer. In either
case, the EM field is able to, in effect, passively ``measure" which
path the particle has followed; in quantum-information-theoretic
language, it can act as a ``witness" as to which path is followed.
This mechanism is of course quite general, and is usually called
``environmental decoherence" \cite{JZ,joos,breuerPet,stamp}; because
of it one expects partial or total destruction of the interference
between the 2 paths.

The particular case of decoherence for 2-path systems in which
electrons couple to photons has been repeatedly discussed in the
literature \cite{imry,ford,breuer}. In this paper we wish to
revisit
this topic, and clarify the role of low frequency infrared (IR)
photons in bringing about the decoherence. Our aim is to unravel the
different
contributions -- from leading and sub-leading soft photon modes -- to
the decoherence. Inevitably, this requires us to discuss how to deal
properly with the infrared divergences present in QED.
At the same time we want to understand how the
leading and sub-leading soft contributions might be observable in
experiments. This turns out to be difficult because they are often
swamped by decoherence from other sources in any realistic experiment.

\subsection{Soft Photons and IR Divergences}

The topic of IR divergences, and their influence on quantum
electronic dynamics, of course has a long history, dating all the way
back to the work of Bloch and Nordsieck \cite{blochN37}. Early work
clarified the ways in which the leading divergences coming from
radiative corrections to scattering processes
\cite{blochN37,sudakov56,abrikosov} were cancelled by divergences
from virtual soft particles, leading to the derivation of
multiplicative  ``soft factors" (see, e.g., refs.
\cite{konoshita50,nakanishi58,murota60}, and also refs.
\cite{yennie61,konoshita62,yennie73,weinberg}) in scattering
processes. These factors, when correctly handled, give finite
multiplicative corrections to scattering cross-sections.

\begin{figure}[ht]
\centering
\includegraphics[scale=0.25]{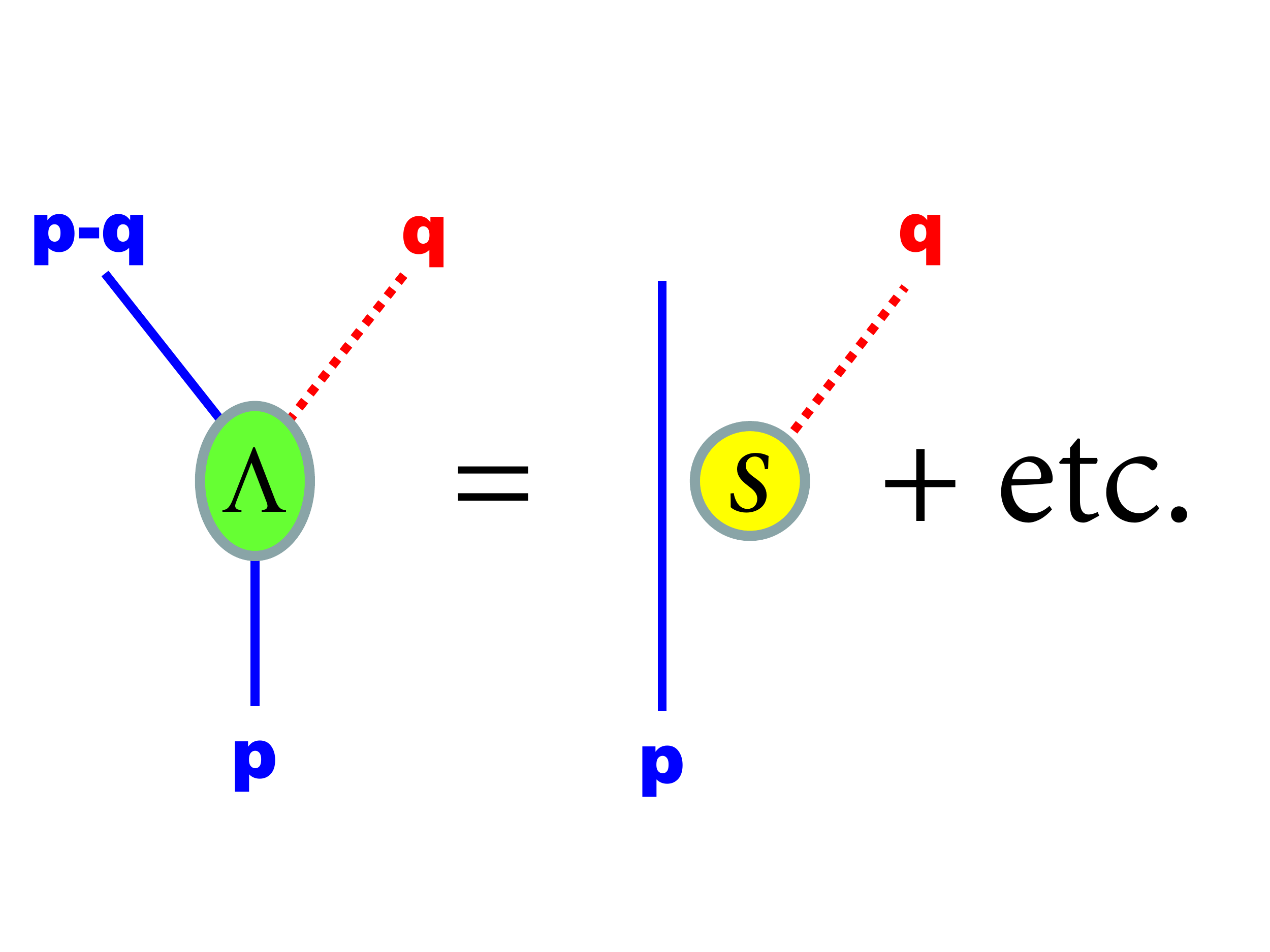}
\vspace{-7mm}
\caption{The 3-point electron-photon vertex $\Lambda(p,q)$ written,
for small $|{\bf q}|$, as a product of the single electron propagator
$K_2(p)$ and the ``soft factor" ${\cal S}(q)$; compare eq.
(\ref{softTh}) in the text. }
 \label{fig:softTh}
\end{figure}

One can write soft factors for any electron-photon scattering
process. Here we are interested in the 3-point electron-photon
scattering process (see Fig. \ref{fig:softTh}), in which the incoming
state is an electron with 4-momentum $p^a$, and the outgoing state is
an electron and a photon, with momenta $p^a-q^a$ and $q^a$
respectively. The ``soft photon theorem'' then says that one can
write the fully renormalized vertex for this process in the form
\begin{equation}
\Lambda(p,q) \;=\; K_2(p) \, \left[{\cal S}_{(0)}(q) + {\cal
S}_{(1)}(q)\right] \;+\;  \mathcal{O}(|{\bf q}|)
 \label{softTh}
\end{equation}
where $K_2(p)$ is the fully renormalized 2-point propagator for the
electron alone, with no photon emission. Here
the divergent leading soft factor ${\cal S}_{(0)}(q) \sim
\mathcal{O}(1/|{\bf q}|)$, where ${\bf q}$ is the 3-momentum, and the
sub-leading factor ${\cal S}_{(1)}(q)) \sim \mathcal{O}(1)$.


Later work has shown that this result can also be understood as a
consequence of asymptotic symmetries \cite{he1,kapec2,lysov}, or by
using either an eikonal analysis \cite{fradkin63,fradkin66,jordan18},
or a coherent state representation for the coupled electron-photon
system \cite{chung,kibble1,kibble2,kibble3,kibble4,FK}.  The latter
representation, in which the system is represented by ``dressed"
electron states, is quite illuminating, for two reasons.

(1) First, dressed states are physically intuitive.
As discussed by e.g. Faddeev and Kulish \cite{FK}, asymptotic
states used in scattering calculations should not be chosen to
represent electrons which are totally decoupled from the Maxwell
field. Since electromagnetic interactions are long-range,
electrons are at no point ever ``undressed.'' One should
instead always work with dressed states, which represent
electrons accompanied by a cloud of low energy photons.
This prescription has consequences when trying to calculate
decoherence: the decoherence rate one finds will generally
diverge if one evaluates scattering processes between the usual
Fock states, but will be convergent if one instead uses dressed
states \cite{gordon1,gordon4}.

(2) Second, the coherent state representation affords a nice way of
separating the leading contributions to the scattering processes from
all sub-leading contributions, first discussed in perturbative
calculations \cite{low1,low2,BK,GG} (this separation is also done in
a very clear way in the eikonal expansion \cite{fradkin63,fradkin66}).

The separation of terms will be crucial in what follows, since one
question we wish to address here is whether IR dressing should be
applied at both leading and sub-leading orders.

One reason for asking this question is that,
while both the leading and sub-leading soft photon theorems have
been related to asymptotic symmetries,
the nature of these symmetries are subtly different.
Namely, the leading order soft photon theorem has been
associated \cite{he1,kapec2} with a group of ``large'' gauge
transformations -- those which have finite, angle-dependent limits
to null infinity. On the other hand, the symmetries associated
with the sub-leading soft photon theorem cannot be
straightforwardly understood as gauge transformations in their
usual form \cite{lysov} (although some progress has been
made on advancing such an interpretation \cite{campiglia}).

So, while dressed states have been constructed even to
sub-leading order \cite{choi}, only the leading order dressing is
obviously necessary to satisfy constraints imposed by
demanding gauge-invariance asymptotically \cite{kapec}.
This difference between leading and sub-leading soft modes
will be mirrored in the calculation that follows. The leading order
dressing will turn out to be necessary to alleviate IR
divergences, while no such divergences ever appear at
sub-leading order.

To show this, and to quantify the leading and sub-leading
contributions to decoherence, we will use the recent discovery
that the leading and sub-leading soft photon factors are
encoded at the endpoints of charged particle worldlines \cite{me}.
This result allows us to cleanly isolate the quantum information
content of both leading and sub-leading soft photons in our model.
What is more, it is not necessary to go to any asymptotic limit to
obtain these results. This is important -- while many questions about
the information content of soft photons are asked in the context of
black holes \cite{hawking16} or scattering processes taking place
over infinite amounts of time \cite{gordon1,gordon4}, we will be
able to come to our conclusions even in the context of a somewhat
idealized interferometry experiment, in which all processes are confined
to a finite space-time region. Thus we have
the possibility of testing for the size of sub-leading contributions
to decoherence, and deciding empirically the question posed above.

Real interferometry experiments, as well as 2-slit
experiments, are much more complicated than the idealization we
introduce. There are multiple other sources of decoherence, including
electronic 4-currents (in conducting systems), phonons, two-level systems
made from electric dipoles or charges hopping between sites,
paramagnetic impurities, and so on; we note that static charges and
electric dipoles can interact over long ranges with the electron.
One can also consider large neutral objects, but these are polarizable and
still interact weakly with long wavelength photons; however they typically
interact rather strongly with other decoherence sources like phonons. Later
in the paper we discuss the experimental prospects in more detail.

\subsection{Summary \& Structure of the Paper}

The rest of the paper proceeds as follows.

First, in Section \ref{sec:model} we introduce what is by now a
standard schematic semiclassical interferometry model. We describe
the assumed geometry of the interferometer, and mention the
various assumptions and approximations we will use.

Section \ref{sec:which_path} introduces the way in which we quantify
the which-path information obtained by photons radiated by the
superposed charge. As we will show, this can be done rather neatly in
terms of the decoherence functional $\Gamma$.

In Section \ref{sec:currents} we discuss relevant features of the
momentum space electromagnetic current $j^a(q)$ which appears in the
decoherence functional. This allows us to review results from our
recent work \cite{me}, which showed that boundary terms in the
electromagnetic current encode the ``soft factors'' appearing in both
the leading and sub-leading soft photon theorems.

We then show in Section \ref{sec:dressing} that superpositions of the
leading soft factor generically cause the decoherence functional
$\Gamma$ to diverge in the infrared -- even for finite-time
experiments. We must deal with these infrared divergences properly in
order to obtain sensible results for the quantum information content
of the photon field, so Section \ref{sec:dressing} also discusses how
we ought to ``dress'' the time evolution operator in our
semiclassical model in order to render the decoherence functional
infrared-finite. After adding the appropriate dressing, the leading
soft photons cease to obtain any quantum information about the matter
particle.

Section \ref{sec:eval} contains the evaluation of the corrected
expression for $\Gamma$ in our model. We show that sub-leading soft
photons may still contribute to $\Gamma$, and we isolate the
contribution coming from the difference in the sub-leading soft
current on each branch of the superposition. The sub-leading
contribution to decoherence turns out to be time-independent, and to
depend only upon the spatial extent of the model interferometer.

Although the inherent IR-finiteness of the
sub-leading contribution suggests that sub-leading dressing is not
necessary, ideally the question of whether or not sub-leading
dressing exists should be answered experimentally.
Accordingly, Section \ref{sec:subleading_dressing} also contains
a calculation of decoherence when a simple form of sub-leading
dressing is applied. The results with and without sub-leading dressing
are sufficiently different that we believe it should be in principle
possible to detect the presence or absence of this extra dressing
in the lab.

In any real experiment however there will be other processes
also causing decoherence -- these include coupling of the electron to
any gas particles in the experiment, the long-range
interaction of the electron with ``charge fluctuators" and electronic
currents in the apparatus. The first can be understood using scattering theory
\cite{scatt}, the second using a ``spin bath" theory
\cite{RPP00}, and the final mechanism using an oscillator bath model for the
electronic environment. We discuss these experimental considerations further
in Section \ref{sec:Expt}.

The paper concludes in Section \ref{sec:discussion} with a discussion
of our results, and of some open questions.

Here we use the mostly-negative metric signature, $\eta_{ab} \equiv
\textrm{diag}(1, -1, -1, -1)$, units in which $\hbar = c = \epsilon_0 =
1$, and a bar over a quantity will indicate its complex conjugate.


\section{Interferometer Model} \label{sec:model}

\begin{figure}[ht]
\centering
\begin{tikzpicture} [scale=2]   

\coordinate (I) at (-.3,.5);
\coordinate (BS_1) at (.5,.5);
\coordinate (M_L) at (.5, 2.5);
\coordinate (M_R) at (2.5,.5);
\coordinate (F) at (2.5,2.5);

\draw [middlearrow={>},thick,double] (I) -- (BS_1);
\draw [middlearrow={>},very thick] (BS_1) -- node[left, scale=1]
{$\dot{X}^a_1$} (M_L);
\draw [middlearrow={>},very thick] [dashed] (BS_1) -- node[above,
scale=1] {$\dot{X}^a_2$} (M_R);
\draw [middlearrow={>},very thick] (M_L) -- node[below, scale=1]
{$\dot{X}^a_2$} (F);
\draw [middlearrow={>},very thick] [dashed] (M_R) -- node[left,
scale=1] {$\dot{X}^a_1$} (F);

\draw [->] (-.5,2) -- (-.1,2) node [right, scale=0.8] (xlabel) {$x$};
\draw [->] (-.5,2) -- (-.5,2.4) node [above, scale=0.8] (ylabel)
{$y$}; 

\draw [<->] ($(BS_1) + (.25,0.08)$) -- node[right, scale=1] {$l$}
($(M_L) + (.25,-.08)$);

\node[circle, draw=teal, fill=lime, scale=.8] (c) at (BS_1){$X^a_i$};
\node[circle, draw=teal, fill=lime, scale=.8] (c) at (M_L){$X^a_L$};
\node[circle, draw=teal, fill=lime, scale=.8] (c) at (M_R){$X^a_R$};

\node [draw=purp_line,
    fill=purp_bg,
    diamond,
    minimum width=1.3cm,
    minimum height=1.3cm
]  (detector) at ($(F) + (0.1,0.1)$){$\mathbf{D}$};

\end{tikzpicture}
\caption{Top-down view of the two-path geometry traversed by the
charged particle.}
 \label{fig:2path}
\end{figure}
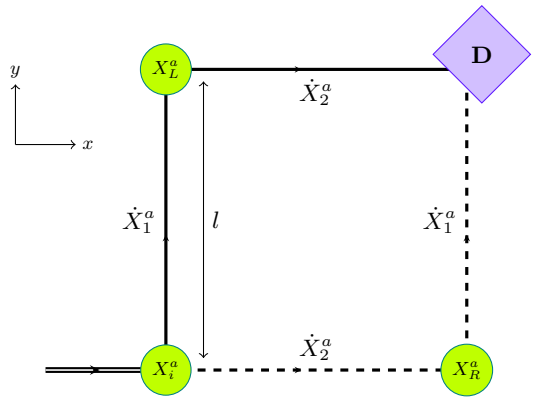

Let us begin by introducing the simple model we will use to describe
a charged particle traversing an interferometer. We emphasize that
although this model is fairly standard in quantum optics \cite{aspect},
it is rather schematic in nature -- in Section 7 we discuss what kind of
elaborations may be required for any ``real world'' treatment of this
system. The spatial trajectories taken by the superposed particle
are illustrated in Figure \ref{fig:2path}. The particle enters the
apparatus along the $\hat{x}$ direction with four-velocity
$\dot{X}^a_2$; upon reaching the point $X^a_1$ at time $t=0$, the
particle is put into a superposition, in which it is either kicked
into the $\hat{y}$ direction to follow the trajectory $X_L^a(s)$,
or allowed to proceed in the $\hat{x}$ direction and follow the
trajectory $X_R^a(s)$. Here $s$ is the proper time experienced
by the particle as it evolves along the trajectories.



As the system then evolves, that branch of the superposition
following the trajectory $X_L^a(s)$ (the solid line in Figure
\ref{fig:2path}) proceeds with the four-velocity $\dot{X}_1^a$ to the
point $X_L^a$, and then with four-velocity $\dot{X}_2^a$ until it
reaches the detector $\mathbf{D}$. The branch following the
trajectory $X_R^a(s)$ (dashed line) proceeds with the four-velocity
$\dot{X}_2^a$ to the point $X_R^a$, and then with four-velocity
$\dot{X}_1^a$ to the detector. On both paths, the speed of the
particle is taken to be constant at $v \equiv l/\tau$, where $l$ is
the length of one side of the interferometer, so that the particle
reaches either $X^a_L$ or $X^a_R$ at time $t = \tau$ and reaches the
detector at $t=2\tau$.

Fig. \ref{fig:2path} is schematic in the sense that in reality, a
{\it single} path shown going along the left or right trajectory
actually represents the set of {\it all} paths passing through the
left arm of an interferometer (this shorthand way of representing
entire classes of path originated in Feynman's celebrated description
of two-path experiments \cite{feynman3}, and can be made more precise
than is necessary here \cite{2class}). It is
assumed in what follows that the experimental setup is ``well-designed",
in the sense that these 2 classes of electronic path
are well-separated in space, and that the electronic acceleration when
following them is very different for each one of them.

In the computations that follow, we are typically interested in
the interaction between the particle and the long-wavelength part of
the electromagnetic field. Therefore we assume, following
\cite{ford,breuer}, that any photons excited during the experiment
will have a wavelength which is too large (much larger than the
particle's de Broglie wavelength) for them to meaningfully resolve
the spread of the particle's wavefunction. The electromagnetic field
in our model then ``sees'' each trajectory as a single path, and so
experiences a simple discrete superposition of two classical
electromagnetic currents, $j^a_L$ and $j^a_R$ in the interferometer.
The assumption that the experimental setup is
well-designed then means that $j^a_L$ and $j^a_R$ are quite distinct, and
so too will be the corresponding photon states.

This leads us to approximate the state of the matter+radiation system
as the particle enters the detector $\mathbf{D}$ using the simple
two-path form
\be \label{eq:final_state}
\frac{1}{\sqrt{2}} \left[ |\psi_L\rangle |L\rangle + |\psi_R\rangle
|R\rangle  \right] ,
\ee
where $|\psi_{L/R}\rangle$ and $|L/R\rangle$ are the states of the
matter particle and the radiation field respectively, after the
particle traverses the $L/R$ arm of the interferometer.

The electromagnetic current of a pointlike particle with charge $e$,
which follows either of the trajectories $X_{L/R}^a(s)$, is
\be \label{eq:Current}
j_{L/R}^a(x) = e\int_{s_i}^{s_f} ds \, \dot{X}_{L/R}^a(s)
\delta^{(4)}(x - X_{L/R}(s) ),
\ee
where again $s$ is the particle's proper time, and
$\dot{X}_{L/R}^a(s)\equiv \frac{d}{ds}X_{L/R}(s)$ its four-velocity.
The bounds on this integral are chosen such that the initial and
final proper times coincide with the beginning and end of the
experiment, $X_{L/R}^0(s_i) = 0$ and $X_{L/R}^0(s_f) = 2\tau$.

Assuming the photon field is in vacuum at the time the experiment
starts, under time evolution the electromagnetic environment will be
placed into a superposition of coherent photon states, sourced by
either the $L$ or $R$ current:
\be \label{eq:coherent_states}
|L/R \rangle \equiv \mathcal{T} e^{-i\int d^4x \; j_{L/R}^a(x)
\hat{A}_a(x)} |0\rangle
\ee
These coherent states are obtained by the action of the
interaction-picture time evolution operator acting on the photon
vacuum $|0\rangle$, and here $\mathcal{T}$ denotes time-ordering. In
writing this expression we have assumed as in \eqref{eq:Current} that
the currents $j_{L/R}^a(x)$ have support only for the duration of the
experiment (between $s_i$ and $s_f$), so we can let the integral
$\int d^4x$ run over all values of $t\in (-\infty, +\infty)$.


\section{MEASURES OF WHICH-PATH INFORMATION AND DECOHERENCE}
\label{sec:which_path}


Our aim here is to quantify the which-path information gained by the
infrared radiation emitted by the charged particle as it proceeds
through the interferometer. To do so, we will make use of two
standard measures of such information, known as
\textit{path distinguishability} and
\textit{interferometric visibility} \cite{WZ, GY, Englert}.

\subsection{Path Distinguishability and Interferometric Visibility}

Suppose that some system of interest $\mathcal{S}$ begins in a pure
state, and then undergoes 2-path interference. Let the state of the
system after traversing solely the first and second paths be called
$|\mathcal{S}_1\rangle$ and $|\mathcal{S}_2\rangle$ respectively.
The state occupied by some which-path measuring device
$\mathcal{M}$ -- also assumed to
start in a pure state -- after the system traverses the first
(second) path, we call $|\mathcal{M}_1\rangle$
($|\mathcal{M}_2\rangle$). Then after the system follows a balanced
superposition of both paths, the combined state of the system and
measuring device will be of the form
\be \label{eq:superposition_state}
|\Psi\rangle = \frac{1}{\sqrt{2}}\left[ |\mathcal{S}_1\rangle
|\mathcal{M}_1\rangle + |\mathcal{S}_2\rangle |\mathcal{M}_2\rangle
\right].
\ee

The path distinguishability $\mathcal{D}$ is defined to be the trace
distance between the final states $\rho^\mathcal{M}_{1/2} \equiv
|\mathcal{M}_{1/2}\rangle\langle \mathcal{M}_{1/2}|$ of the
measurement apparatus $\mathcal{M}$ on each branch of the
superposition:
\be \label{eq:d_defn}
\mathcal{D} \equiv \frac{1}{2} \textrm{Tr} |\rho^\mathcal{M}_{1}  -
\rho^\mathcal{M}_{2} | = \sqrt{1 - |\langle \mathcal{M}_{2} |
\mathcal{M}_{1} \rangle|^2 }
\ee
This path distinguishability yields an upper bound to the
likelihood $\mathcal{L}$ of
successfully discriminating between the two paths taken by the system
$\mathcal{S}$ by observing only the degrees of freedom of the
measurement device $\mathcal{M}$, as $\mathcal{L} \le \frac{1}{2}(1 +
\mathcal{D})$ \cite{Englert}, making it the natural measure of how
much which-path information is contained in $\mathcal{M}$.

The interferometric visibility $\mathcal{V}$, on the other hand,
quantifies how easy it is to observe the coherence of the system
$\mathcal{S}$. $\mathcal{V}$ is typically defined in the context of
double-slit interferometry as
\be
\mathcal{V} \equiv \frac{I_{max} - I_{min}}{I_{max} + I_{min}},
\ee
where $I_{max}$ and $I_{min}$ are the maximum and minimum intensities
of the interference fringes on a screen. More generally it obeys the
upper bound
\be \label{eq:v_upper_bound}
\mathcal{V} \le |\langle \mathcal{M}_2 | \mathcal{M}_1 \rangle|,
\ee
though its precise form depends upon the specifics of the
interferometer (see for example \cite{QV} for an explicit calculation
of $\mathcal{V}$ in a double-slit interferometer).

Eqs. \eqref{eq:d_defn} and \eqref{eq:v_upper_bound} lead immediately
to the well-known ``duality relation'' between distinguishability and
visibility,
\be
\mathcal{D}^2 + \mathcal{V}^2 \le 1.
\ee
This relation expresses the fact that the acquisition of quantum
information is generally destructive; the more which-path information
is obtained by $\mathcal{M}$, the less coherent $\mathcal{S}$ will
appear.

\subsection{The Electromagnetic Field as a Measuring Device}

Let us now return to the charged particle introduced in the previous
section. Clearly the combined state in eq. \eqref{eq:final_state}
has the same form as \eqref{eq:superposition_state}, with the role
of $\mathcal{S}$ played by the moving charge, and $\mathcal{M}$
played by the photon field. Quantifying the amount of which-path
information carried away via photon emission, and the resulting loss
of interferometric visibility, then follows immediately. The
which-path information gained by the photons is
\be
\mathcal{D} = \sqrt{1 - |\langle R | L \rangle|^2},
\ee
and the interferometric visibility in an experiment of the form we
consider is bounded from above according to
\be \label{eq:v_rl}
\mathcal{V} \le |\langle R | L \rangle|.
\ee

In our toy model interferometry setup the quantity governing both the
visibility $\mathcal{V}$ and the distinguishability $\mathcal{D}$ --
and
therefore how much which-path information is obtained by the
electromagnetic field -- is then the modulus of the inner product
between the photon coherent states, which we write in the form
\be \label{eq:LR_inner_product}
|\langle R | L \rangle| \equiv e^{- \Gamma}; \hspace{1em} \Gamma \in
\mathbb{R} , \hspace{1em} \Gamma \ge 0.
\ee
The quantity $\Gamma$ appearing is just
the well-known {\it decoherence functional}
\cite{imry,breuerPet,decoF,feynmanV63}, so called because it is a
functional of the currents along the entire $L$ and $R$ paths, ie.,
$\Gamma \equiv \Gamma[j_L, j_R]$. It thus depends on everything that
happens
to the charge on each of these paths, and neatly encodes the process of
emission of which-path information into the electromagnetic
environment.

In the case that the decoherence is very small -- and we do
expect the contribution to decoherence coming from radiation of
infrared photons to be small in laboratory settings -- we can
approximate $|\langle R | L \rangle| \approx 1 - \Gamma$. Then we
have
\be
\mathcal{D} \approx \Gamma; \hspace{2em} \mathcal{V} \lesssim 1 -
\Gamma.
\ee
The decoherence functional and the path distinguishability are
equivalent in the context of our model interferometry experiment as
measures of how much which-path information is obtained by the
radiated photons, and $\Gamma$ also directly quantifies the loss of
visibility due solely to photon emission. Because $\Gamma$ so cleanly
captures the tradeoff between $\mathcal{D}$ and $\mathcal{V}$,
we will work solely with $\Gamma$ in what follows.


\section{Form of Electromagnetic Currents \& Decoherence Functional}
\label{sec:currents}


The electromagnetic decoherence functional was first derived long ago
\cite{feynmanV63,FeynmanHibbs}, and has been discussed more recently
e.g. by Ford \cite{ford}, and by Breuer and Petruccione
\cite{breuer}. At zero temperature it takes the form
\be \label{eq:gamma_m}
\Gamma[j_L, j_R] =   {1\over2}\int \frac{d^3q}{(2\pi)^3 2 \omega} \;
P_{ab} \, \delta  j^a(q) \delta  j^b(-q),
\ee
where $\delta j^a \equiv j_L^a - j_R^a$ is the difference in
electromagnetic currents along the two superposed paths, $q^a \equiv
\omega(1,\hat n)$, with $\hat n$ an angular unit vector, is a null
four-momentum, and $\int \frac{d^3q}{(2\pi)^3} \equiv \oint
{dS^2(\hat n)\over (2\pi)^2} \int_0^\infty { d\omega \over 2\pi}
\omega^2 $. Finally, the transverse projector $P_{ab}$ can be written
in terms of the momentum $q^a$ as $P_{jk} \equiv \delta_{jk} -
\frac{q_j q_k}{\omega^2}$, with $P_{0a} = 0$.

Note that $\Gamma$ as written is a functional of currents which may
have support for times $t\in(-\infty, +\infty)$. However we have
restricted to finite-time dynamics here by restricting the range of
the proper time integrals in \eqref{eq:Current}.

Let us now further discuss the matter currents we will insert into
the decoherence functional \eqref{eq:gamma_m}; we will wish to
understand in particular the various contributions to the decoherence
functional -- which is central to any analysis of decoherence, for
any system -- coming from the endpoints of the particle's paths
through the interferometer.

What appears in the decoherence functional is actually the Fourier
transform of the current, $j^a(q) \equiv \int d^4x \, e^{iq\cdot x}
j^a(x)$, which using \eqref{eq:Current} is
\be
\label{eq:j_first_expression}
j^a(q) \;=\;  e \int ds\,\dot{X}^a(s) \, e^{iq\cdot X(s)} .
\ee
which we rewrite as
\be
j^a(q) \;=\;  e \int ds\, \dot{X}^a \left( \frac{1}{iq\cdot \dot{X}}
\right) \frac{d}{ds} e^{iq\cdot X} ,
\ee
so that integrating by parts in $s$ gives
\be \label{eq:j_with_bdry_term}
\begin{split}
j^a(q) = &-ie \int ds\,\frac{d}{ds} \left(  e^{iq\cdot X}
\frac{\dot{X}^a}{q\cdot \dot{X}}  \right) \\
&+ ie \int ds\, e^{iq\cdot X} \frac{d}{ds} \left(
\frac{\dot{X}^a}{q\cdot \dot{X}} \right) .
\end{split}
\ee
Discarding the boundary term appearing here
\cite{ilderton,laddha1,me} ensures that the current is conserved,
$q_a j^a(q) = 0$, leaving us with the following corrected expression
for the particle current in momentum space:
\be \label{eq:jq_final}
j^a(q) =  ie \int ds\, e^{iq\cdot X} \frac{d}{ds} \left(
\frac{\dot{X}^a}{q\cdot \dot{X}} \right)
\ee

Notice that the form \eqref{eq:jq_final} of the matter current has the
property that the current vanishes when the particle is not
accelerating ($\ddot{X}^a = 0$).

Plugging the current into the decoherence
functional
\eqref{eq:gamma_m}, we note immediately the important fact that
\textit{there is no decoherence without acceleration}, because the
decoherence in this system arises purely as a result of of radiative
coupling to the photon field.

\subsection{Soft Electromagnetic Currents}

Previously we showed \cite{me} that the electromagnetic current of a
point particle \eqref{eq:jq_final} is the sum of two terms, each of
which  which are entirely localized on the boundary (ie., the end
points)
of the worldline the particle traces out in space-time. These boundary
terms turn out  to be those parts of the current which dominate at low
frequencies.

In particular we showed that, expanding \eqref{eq:jq_final} in powers
of $q$, we get
\begin{eqnarray}
 \label{eq:Jboundary}
ij^a(-q) \;&=&\;  e \left[ \Delta [\mathcal{S}^a_{(0)}(q, m\dot{X})] +
\Delta [\mathcal{S}^a_{(1)}(q, X, m\dot{X})] \right] \nonumber \\
&& \qquad\qquad + \mathcal{O}(q).
\end{eqnarray}
where, if the proper time integral in \eqref{eq:jq_final} runs from $s_i$ to
$s_f$, then the $\Delta[\cdots]$ in this expression has the meaning
\be
\Delta [f(s)] \equiv f(s_f) - f(s_i) = \int_{s_i}^{s_f} \partial_s
f(s),
\ee
showing that these contributions live on the endpoints of the
particle worldline, at $s_i$ and $s_f$, and we have defined the
leading and sub-leading electromagnetic ``soft factors''
\be \label{eq:soft_factor_defn}
\mathcal{S}_{(0)}^a(q,p) \equiv   \frac{p^a}{q\cdot p} ,\hspace{1em}
\mathcal{S}_{(1)}^a(q,x,p) \equiv  i \frac{q_b{J}^{ba}}{q\cdot p} .
\ee
Here, ${J}^{ab} \equiv 2 p^{[a}x^{b]}$ is the angular momentum
tensor. Ref. \cite{me} further showed that these boundary terms
explain the factorization of the leading and sub-leading soft photon
theorems at tree level.

We emphasize that these boundary terms on the worldline are
\textit{not} necessarily located at asymptotic coordinate times
$t\rightarrow\pm\infty$. They exist and take the form
\eqref{eq:Jboundary} for any choice of bounds on the proper time
integral in \eqref{eq:jq_final}.

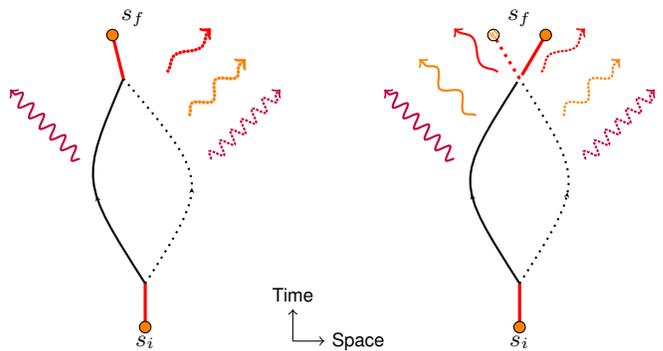
\begin{figure}[t!]
\centering
    \begin{tikzpicture} [scale=2.9]   

    \coordinate (I) at (0,-.67);
    \coordinate (F) at (-.15,.67);

    \draw [thick, rounded corners, middlearrow={>}] ($(I) + (0,0.2)$)
    .. controls (-.3,0)  .. ($(F) - (-.05,0.2)$);
    \draw [thick, rounded corners, dotted, middlearrow={>}] ($(I) +
    (0,0.2)$) .. controls (.3,0)  .. ($(F) - (-.05,0.2)$) ;

    \draw[snake it,segment aspect=0, segment length=2mm,
    purple,thick,->] (-.3,.1) -- (-.62,.42);
    \draw[snake it,segment aspect=0,purple, segment length=2mm,
    thick,dash pattern=on 1pt off .5pt,->] (.3,.1) -- (.62,.42);

    \draw[snake it,segment aspect=0, segment length=4.5mm,
    orange,very thick,dash pattern=on 1pt off .5pt,->] (.2,.3) --
    (.46,.56);
    \draw[snake it,segment aspect=0, segment length=4.5mm, orange,->]
    (.2,.3) -- (.46,.56);

    \draw[snake it,segment aspect=0, segment length=7mm, red,very
    thick,dash pattern=on 1pt off .5pt,->] (.1,.5) -- (.3,.7);
    \draw[snake it,segment aspect=0, segment length=7mm, red,->]
    (.1,.5) -- (.3,.7);

    \draw [very thick, rounded corners, red] (I) -- ($(I) +
    (0,0.2)$);
    \draw [very thick, rounded corners, red]  ($(F) - (-.05,0.2)$) --
    (F);

    \node [circle, draw=black, fill=orange, scale=0.5] at (I) {};
    \node [below] at (I) {$s_i$};
    \node [circle, draw=black, fill=orange, scale=0.5] at (F) {};
    \node [above right] at (F) {$s_f$};

    \end{tikzpicture}\hspace{-1em}
    \begin{tikzpicture}[scale=1.2]
    \scriptsize
    \draw [->] (.4,-.6) -- (.4,-.25) node [above] (timelabel) {Time};
    \draw [->] (.4,-.6) -- (.75,-.6) node [right] (spacelabel)
    {Space}; 
    \end{tikzpicture} \hspace{-1em}
    \begin{tikzpicture} [scale=2.9]   

    \coordinate (L) at (-2,0);
    \coordinate (R) at (2,0);

    \coordinate (B) at (0,-2);

    \coordinate (I) at (0,-.67);
    \coordinate (F) at (0,.67);
    \coordinate (FL) at (-.12,.67);
    \coordinate (FR) at (.12,.67);

    \coordinate (T) at ($(F) - (0,0.2)$);

    \draw [thick, rounded corners, middlearrow={>}] ($(I) + (0,0.2)$)
    .. controls (-.3,0)  .. ($(T) - (0.01,0.015)$);
    \draw [thick, rounded corners, dotted, middlearrow={>}] ($(I) +
    (0,0.2)$) .. controls (.3,0)  .. (T) ;

    \draw[snake it,segment aspect=0, segment length=2mm,
    purple,thick,->] (-.3,.1) -- (-.62,.42);
    \draw[snake it,segment aspect=0,purple, segment length=2mm,
    thick,dash pattern=on 1pt off .5pt,->] (.3,.1) -- (.62,.42);

    \draw[snake it,segment aspect=0, segment length=4.5mm, orange,
    thick,dash pattern=on 1pt off .5pt,->] (.2,.3) -- (.46,.56);
    \draw[snake it,segment aspect=0, segment length=4.5mm, thick,
    orange,->] (-.2,.3) -- (-.46,.56);

    \draw[snake it,segment aspect=0, segment length=7mm, red,
    thick,dash pattern=on 1pt off .5pt,->] (.1,.5) -- (.3,.7);
    \draw[snake it,segment aspect=0, segment length=7mm, thick,
    red,->] (-.1,.5) -- (-.3,.7);

    \draw [very thick, rounded corners, red] (I) -- ($(I) +
    (0,0.2)$);
    \draw [very thick, rounded corners, dotted, red]  (T) -- (FL);
    \draw [very thick, rounded corners, red]  ($(T) +
    (0.011,0.0165)$) -- (FR);

    \node [circle, draw=black, fill=orange, scale=0.5] at (I) {};
    \node [below] at (I) {$s_i$};
    \node [circle, draw=black, fill=orange, scale=0.5] at (FR) {};
    \node [circle, draw=black,
    pattern={Lines[angle=-57,distance=1pt]}, pattern color=orange,
    scale=0.5] at (FL) {};
    \node [above] at (F) {$s_f$};

    \end{tikzpicture}

    \caption{Left -- two charged particle trajectories which source
    different hard radiation, but the same soft radiation. Right --
    two trajectories which source different hard and soft radiation,
    due to differences in the soft currents $j_{div}^a$ and
    $j_{sub}^a$ at the worldline boundary. The hard radiation is
    shown as the shortest wavelength emission, in purple; the leading
    soft radiation shown as the longest wavelength emission, in red;
    and the sub-leading soft radiation in orange.}
    \label{fig:currents}
\end{figure}

From eq. \eqref{eq:Jboundary} it is then clear that we can always
split the point particle current into three pieces, viz.,
\be \label{eq:soft_current}
j^a(q) \equiv j_{div}^a(q) + j_{sub}^a(q) + j_{hard}^a(q),
\ee
with
\be \label{eq:soft_current_parts}
\begin{split}
j_{div}^a(q) &\equiv ie \Delta \left[\frac{\dot{X}^a}{q\cdot \dot{X}}\right]  \\
j_{sub}^a(q) &\equiv  e \Delta \left[\frac{q_b[ X^a \dot{X}^b  - \dot{X}^a
X^b ]}{q \cdot \dot{X}}\right] \\
j_{hard}^a(q) &\equiv j^a(q) - j_{div}^a(q) - j_{sub}^a(q) .
\end{split}
\ee
We interpret the currents $j_{div}^a$, $j_{sub}^a$, and $j_{hard}^a$
as the sources of leading soft radiation, sub-leading soft radiation,
and hard radiation, respectively. In this work, we then
\textit{define} ``leading soft radiation'' as radiation which is
sourced by $j_{div}^a$,  ``sub-leading soft radiation'' as radiation
which is sourced by $j_{sub}^a$, and  ``hard radiation'' as radiation
which is sourced by the remaining term $j_{hard}^a$. This
interpretation
is illustrated in Fig. \ref{fig:currents}.

We note immediately that the leading soft current $j^a_{div}$ is
\textit{infrared divergent}, as it goes like $1/\omega$ and blows up
as $\omega\rightarrow0$. The sub-leading soft current $j_{sub}^a$ is
however $\mathcal{O}(1)$ and perfectly infrared-finite, and the hard
current $j_{hard}^a$ is $\mathcal{O}(\omega)$ and vanishes completely
in the soft $\omega\rightarrow0$ limit.  Additionally we note that
each of these currents is individually conserved, i.e.,
\be
q_a j_{div}^a(q) = q_a j_{sub}^a(q) = q_a j_{hard}^a(q) = 0 .
\ee

The identification of these soft currents will allow us to rather
directly assess the contribution to the decoherence functional
\eqref{eq:gamma_m} from leading and sub-leading soft photon modes. In
particular we will expand the decoherence functional
\eqref{eq:gamma_m} in terms of the splitting \eqref{eq:soft_current}
to get
\be
 \label{eq:softDeco}
\begin{split}
\Gamma =   {e^2\over2} &\int \frac{d^3q}{(2\pi)^3 2 \omega} \; P_{ab}
\times \\
& \left[ \delta j_{div}^a(q) + \delta j_{sub}^a(q) + \delta
j_{hard}^a(q) \right] \times \\
& \left[ \delta j_{div}^b(-q) + \delta j_{sub}^b(-q) + \delta
j_{hard}^b(-q)  \right] ,
\end{split}
\ee
displaying clearly the contributions from the leading and sub-leading
soft currents to $\Gamma$.

We will see later on that this expression for $\Gamma$ is unphysical,
since it contains divergent terms that in the end make no
contribution. The true physical form for $\Gamma$ will be  given in
eq. (\ref{eq:gamma_eval_1}) below.


\subsection{Decoherence Functional for Interferometer}

The general form of a decoherence functional $\Gamma$ is independent
of the geometry of the system -- but we will now need the
specific form for our 2-path system.

To obtain this, we first write the electromagnetic currents along the
left and right trajectories as
\be
\begin{split}
j^a_L(q) = ie \int ds \, e^{i q \cdot X_L(s)}  \partial_s
\left[\frac{\dot{X}_L^a(s)}{q \cdot \dot{X}_L(s)}\right]  , \\
j^a_R(q) = ie \int ds \, e^{i q\cdot X_R(s)}  \partial_s
\left[\frac{\dot{X}_R^a(s)}{q\cdot \dot{X}_R(s)}\right]  .
\end{split}
\ee
The difference between the two currents on the superposed paths,
\be
\delta j^a(q) \equiv j^a_L(q) - j^a_R(q)
\ee
can then be approximated in the geometry illustrated in Figure
\ref{fig:2path} as
\be \label{eq:dj1}
\delta j^a(q) = ie \,  \left[ e^{iq \cdot X_i} - e^{iq\cdot X_L} -
e^{iq \cdot X_R }  \right]
\left[\frac{\dot{X}_1^a}{q\cdot \dot{X}_1} -
\frac{\dot{X}_2^a}{q\cdot \dot{X}_2}\right] .
\ee
Notice that the current difference consists of terms associated with
the points $X_i^a, X_L^a,$ and $X_R^a$. These are the space-time
events at which the particle accelerates and thence radiates photons.

Expressions like (\ref{eq:softDeco}) and (\ref{eq:dj1}) can only be
used once the trajectories for the electron are specified. This can
only be done in detail for a specific experimental setup - for which
see section 7. In what follows we will simply assume

(i) that the particle travels non-relativistically -- by which we
mean that its velocity $v$ satisfies
\be
v \equiv {l\over \tau} \ll 1
\ee
in units with $c=1$ -- then we can assume that the spatial parts of
the phases appearing in \eqref{eq:dj1} can be neglected to leading
order. This is just the well-known dipole approximation
\cite{jackson}; we are assuming, as noted in the introduction, that
the radiation emitted during the experiment will have a wavelength
that is long compared to the spatial extent of the interferometer,
and so cannot resolve the spatial separation of the points at which
the particle radiates. We will also assume
(ii) that the spatial extent of the ``scattering'' regions $X^a_{\{ i, L,
R\}}$ inside of which the particle experiences acceleration, can be
characterized by a length scale $l_{0} \equiv 1 / \Omega$, where
$\Omega$ effectively acts as a UV frequency cutoff for the scattering.
It follows that $l_0$ is then the shortest length scale relevant to
our analysis. An ultraviolet cutoff $\Omega$ of this sort also emerges
naturally if one smears out the superposed trajectories we are considering,
allowing the currents to have support along worldtubes of finite width $\sim
1/\Omega$, rather than on perfectly-localized one-dimensional
worldlines \cite{breuer}.

Note that the timescale $\tau$ over which the experiment occurs is then
long, in the sense that $\Omega \tau \gg 1$, because $v \ll 1$. To put it
another way, the timescale $\tau$ is much greater than the time it takes
light to cross the scattering regions, of size $1/\Omega$.

With these assumptions in place, the current difference can be further
simplified to
\be \label{eq:dj_approx}
\delta j^a(q) \approx ie  \left( 1 - 2 e^{i\omega\tau} \right)
\left[\frac{\dot{X}_1^a}{q\cdot \dot{X}_1} -
\frac{\dot{X}_2^a}{q\cdot \dot{X}_2}\right]  .
\ee

Plugging this current difference into the decoherence functional
gives
\begin{align} \label{eq:fullgamma}
\Gamma &\approx {e^2 \over 4(2\pi)^3} \int_\lambda^\Omega
{d\omega\over\omega}  \, \left(5 -  4\cos{\omega\tau}  \right) \times
\\
&\oint dS^2(\hat n) \, \omega^2 \, \left[ 2 \frac{\dot{X}_1\cdot
\dot{X}_2}{(q \cdot \dot{X}_1)(q \cdot \dot{X}_2)} - \frac{1}{(q
\cdot \dot{X}_1)^2} - \frac{1}{(q\cdot \dot{X}_2)^2} \right] .
\nonumber
\end{align}
The spherical integral here is independent of $\omega$, since
the term in square brackets goes like ${1 \over \omega^2}$ and this
gets multiplied by the factor $\omega^2$ from the integration
measure.

The result (\ref{eq:fullgamma}) for $\Gamma$ incorporates all the
soft contributions -- what we will now show is that the contribution
from the leading divergent contributions is actually zero.


\section{leading order Soft Photons} \label{sec:dressing}


In order for the integral \eqref{eq:fullgamma} to be well-defined, we
have also been forced to introduce an infrared cutoff $\lambda$. The
integrand of \eqref{eq:fullgamma} diverges like $1/\omega$ as $\omega
\rightarrow 0$, so the frequency integral goes to infinity like $-\ln
\lambda$ as we take the infrared cutoff $\lambda\rightarrow 0$. Taken
na\"ively, this leads one to predict an infinite amount of
decoherence, since $\Gamma \rightarrow \infty$ in the IR limit; this
is clearly unphysical.

In earlier discussion of IR divergences, noted in the introduction,
it was necessary to introduce complicated devices (e.g., measuring
systems designed to observe IR photons) to deal with these
divergences, and show how they were either cancelled or otherwise
eliminated. However in what follows we will show how our
understanding of soft currents allows these divergences to be handled
in a fairly straightforward way, which simply reduces their
contribution to $\Gamma$ to zero.

\subsection{Infrared Divergences}

In section \ref{sec:currents}, we saw that the divergence in
question is due to a superposition of the leading contribution to the
soft current; $\delta j_{div}^a$ is not zero in this example.
Begin by splitting the current \eqref{eq:dj_approx} into its relevant
soft ($\mathcal{O}(\omega^{-1})$ and $\mathcal{O}(1)$) and hard
($\mathcal{O}(\omega)$) pieces. In this model we have $\delta j^a
\equiv \delta j_{div}^a + \delta j_{sub}^a + \delta j_{hard}^a$, with
\begin{align} \label{eq:dj_parts}
\delta j_{div}^a(q) &\approx -ie  \left[\frac{\dot{X}_1^a}{q\cdot
\dot{X}_1} - \frac{\dot{X}_2^a}{q\cdot \dot{X}_2}\right]  \nonumber
\\
\delta j_{sub}^a(q) &\approx 2e\omega\tau
\left[\frac{\dot{X}_1^a}{q\cdot \dot{X}_1} -
\frac{\dot{X}_2^a}{q\cdot \dot{X}_2}\right]  \\
\delta j_{hard}^a(q) &\approx 2ie  \left( 1 -  e^{i\omega\tau}
+i\omega\tau \right) \left[\frac{\dot{X}_1^a}{q\cdot \dot{X}_1} -
\frac{\dot{X}_2^a}{q\cdot \dot{X}_2}\right]  \nonumber
\end{align}
in the dipole approximation.

From the general form of the decoherence functional, eq.
\eqref{eq:softDeco}, we see that $\Gamma$ contains a term involving
only the leading soft current $j_{div}^a$:
\begin{align}
&\Gamma \supset   {1\over2}  \int \frac{d^3q}{(2\pi)^3 2\omega}\;
P_{ab} \, \delta  j_{div}^a(q) \delta  j_{div}^b(-q) \nonumber \\
&= {e^2 \over 4(2\pi)^3} \int_\lambda^\Omega {d\omega\over\omega}
\times \\
&\oint dS^2(\hat n) \, \omega^2 \, \left[ 2 \frac{\dot{X}_1\cdot
\dot{X}_2}{(q \cdot \dot{X}_1)(q \cdot \dot{X}_2)} - \frac{1}{(q
\cdot \dot{X}_1)^2} - \frac{1}{(q\cdot \dot{X}_2)^2} \right] .
\nonumber
\end{align}

It is this term that causes the infrared divergence when taking
$\lambda\rightarrow 0$, seen here as the logarithmically divergent
integral $\int_\lambda^\Omega {d\omega\over\omega}  $. The infrared
divergence is caused by a superposition of the leading soft current,
i.e. $\Gamma \rightarrow \infty$ because $\delta j_{div}^a \neq 0$.
The resulting infinite amount of decoherence should thus be
understood as coming from information loss into leading soft photon
modes.

\subsection{Infrared Dressing}

If this result turned out to be the end of the story, then the
leading order soft photons would always make a perfect which-path
measurement of the charged particle in any experiment of the sort
shown in Figure \ref{fig:2path}. This is obviously wrong: if it were
the case, we would have $\Gamma \rightarrow \infty$ for \textit{any}
choice of experimental parameters ($e, l, \tau$), and from eqs.
\eqref{eq:v_rl} and \eqref{eq:LR_inner_product} we see that there
would \textit{never} be any observable interference once the particle
reaches the detector! Because quantum interference is observed in
Nature, we must reject this result.

This conclusion is not new, and indeed has informed all
investigations of IR divergences in QED since Bloch and Nordsieck
\cite{blochN37}. That one needs to properly handle these divergences
is also obvious in the eikonal formulation of QED
\cite{fradkin63,fradkin66} and was a motivating factor in the
coherent state formulation of QED in terms of dressed states
\cite{chung,kibble1,kibble2,kibble3,kibble4,FK,gordon1}.
In the coherent state formulation,  the IR divergences are eliminated
by dressing the electron, so that soft photons appear already in the
time evolution operator.

It turns out to be incredibly simple to effect IR dressing in our
treatment. We show in Appendix \ref{app:dressing} that the net effect
of applying a ``minimal'' implementation of the dressed formalism,
which we define in that appendix, is that we can simply set the
divergent soft current $j_{div}^a$ to zero everywhere in our
calculations -- the physical result of IR dressing can then be
interpreted as the total decoupling of $j_{div}^a$ from the Maxwell
field.
After incorporating the dressing, when studying the electromagnetic
radiation caused by the charged particle moving through our model
interferometer, what we do is define \textit{dressed coherent states}
in the same way as we defined the coherent radiation states in eq.
\eqref{eq:coherent_states}, but now we explicitly subtract out the
divergent current $j^a_{div}$, i.e., we write:
\be
||L/R \rangle\rangle \equiv  \mathcal{T} e^{-i \int d^4x  \;
[j_{L/R}^a - j_{L/R,div}^a](x) \hat{A}_a(x)} |0\rangle
\ee

We can then define a \textit{dressed decoherence functional} in the
same way we defined $\Gamma$ in the ``undressed'' case, but using the
dressed coherent states instead,
\be
|\langle\langle R | L \rangle\rangle| \equiv e^{-\Gamma_{dressed}} .
\ee
The dressed decoherence functional $\Gamma_{dressed}$ can be obtained
from the undressed decoherence functional \eqref{eq:softDeco} by
simply setting the divergent current $j_{div}^a$ to zero, giving
\be
 \label{eq:gamma_dressed}
\begin{split}
\Gamma_{dressed} =  {e^2\over2} &\int \frac{d^3q}{(2\pi)^3 2 \omega}
\; P_{ab} \times \\
& \left[  \delta j_{sub}^a(q) + \delta j_{hard}^a(q) \right] \times
\\
& \left[ \delta j_{sub}^b(-q) + \delta j_{hard}^b(-q)  \right] .
\end{split}
\ee

The expression \eqref{eq:gamma_dressed} makes it clear that in the
dressed formalism, the leading soft current no longer plays a role in
decoherence of the matter state. Because infrared divergences
necessitate this dressing, we conclude that leading soft photons do
not cause decoherence, and therefore do not carry any quantum
information about the matter.


\section{Sub-Leading Soft Photons} \label{sec:eval}


After having ``dressed away'' the leading soft photon contributions
to decoherence, we can now begin our study of the remaining
sub-leading contributions. We will do this by  once more evaluating
the decoherence functional, but this time using the dressed
expression \eqref{eq:gamma_dressed}.

\subsection{Dressed Decoherence Functional}

We begin by inserting the sub-leading and hard parts of the current
difference from \eqref{eq:dj_parts} into \eqref{eq:gamma_dressed} to
give
\begin{align}
 \label{eq:gamma_eval_1}
&\Gamma_{dressed} \approx {2e^2 \over (2\pi)^3} \int_0^\Omega
{d\omega\over\omega}  \, \left(1 -  \cos{\omega\tau}  \right) \times
\\
&\oint dS^2(\hat n) \, \omega^2 \, \left[ 2 \frac{\dot{X}_1\cdot
\dot{X}_2}{(q \cdot \dot{X}_1)(q \cdot \dot{X}_2)} - \frac{1}{(q
\cdot \dot{X}_1)^2} - \frac{1}{(q\cdot \dot{X}_2)^2} \right] .
\nonumber
\end{align}
This expression incorporates both the sub-leading current difference
$ \delta j_{sub}^a(q)$, and the hard current difference $ \delta
j_{hard}^a(q)$; and cross-terms between them. We stress that since
the contribution form the divergent leading photon terms to $\Gamma$
is zero, eq. (\ref{eq:gamma_eval_1}) should be regarded as the
correct expression for the decoherence functional.

We see that after incorporating the infrared dressing, the integrand
of the frequency integral now goes to zero as $\omega\rightarrow0$,
killing off the divergence we found before, and rendering the
integral finite so that we can evaluate it properly, even with the
infrared cutoff $\lambda$ now taken to zero. The dressing is working
as intended.

To arrive at \eqref{eq:gamma_eval_1}, we have again assumed that the
particle travels through the interferometer non-relativistically,
$v \ll 1$; as discussed earlier, this allowed us to invoke the dipole
approximation, in which we ignored spatial parts of the phases
appearing in the decoherence functional. When evaluating
$\Gamma_{dressed}$ we can again assume that $\Omega \tau \gg 1$, i.e.
that the timescale $\tau$ of the experiment, $\tau$, is such
that $\tau \gg 1/\Omega$.

We evaluate the expression \eqref{eq:gamma_eval_1} using these
approximations in Appendix \ref{app:integrals_dressed}, with the
result that the result for decoherence from the leading
contributions is
\be \label{eq:dressed_result}
\Gamma_{dressed} \approx {4e^2v^2\over3\pi^2} \ln\Omega\tau .
\ee
which agrees with previous calculations of electromagnetic
decoherence \cite{breuer,dominik}.

As expected, this decoherence is coming from dipolar
electromagnetic radiation -- we can write $\Gamma_{dressed} \propto
(\partial_t e\delta x)^2$, where $e\delta x$ is the difference in
electric dipole moments between the $L$ and $R$ branches.

\subsection{Purely Sub-Leading Contribution}

In addition to computing the full amount of photon decoherence in our
model, let us now take a short detour to compute the decoherence
coming only from the sub-leading terms in our dressed expression
(\ref{eq:gamma_eval_1}) for $\Gamma_{dressed}$.

In the general form \eqref{eq:gamma_dressed} of the dressed
decoherence functional, we see that the purely sub-leading term
$P_{ab} \delta j_{sub}^a \delta j_{sub}^b$,  the cross terms $P_{ab}
\delta j_{sub}^a \delta j_{hard}^b$, and the purely hard part $P_{ab}
\delta j_{hard}^a \delta j_{hard}^b$ are all of different orders in
$\omega \tau$, and so can be classified according to this order. The
first purely sub-leading term gives decoherence caused purely by
superpositions of the sub-leading soft current. Writing this
contribution as $\Gamma_{sub}$, we then have
\begin{align}
 \label{eq:sub_gamma}
&\Gamma_{sub} \equiv   {1\over2}  \int \frac{d^3q}{(2\pi)^3
2\omega}\; P_{ab} \, \delta  j_{sub}^a(q) \delta  j_{sub}^b(-q)
\nonumber \\
&= {e^2 \over (2\pi)^3} \int_0^\Omega {d\omega} \; \omega \tau^2
\times \\
&\oint dS^2(\hat n) \, \omega^2 \, \left[ 2 \frac{\dot{X}_1\cdot
\dot{X}_2}{(q \cdot \dot{X}_1)(q \cdot \dot{X}_2)} - \frac{1}{(q
\cdot \dot{X}_1)^2} - \frac{1}{(q\cdot \dot{X}_2)^2} \right] ,
\nonumber
\end{align}
where we have used the expression for $\delta j_{sub}^a$ found in eq.
\eqref{eq:dj_parts}, and where we see that this result is nothing but
the lowest order term in the expansion of the integrand of
(\ref{eq:gamma_eval_1}) in powers of $\omega \tau$.

The quantity $\Gamma_{sub}$ has an intuitive interpretation. Imagine
allowing the electromagnetic field to couple \textit{only to the
sub-leading soft current} $j_{sub}^a$ during our hypothetical
interferometry experiment. The resulting states of the
electromagnetic radiation sourced by the sub-leading current on the
$L/R$ branches would then be
\be
|| l/r \rangle\rangle \equiv  \mathcal{T} e^{-i \int d^4x  \;
j_{L/R, sub}^a(x) \hat{A}_a(x)} |0\rangle ,
\ee
and $\Gamma_{sub}$ would encode the modulus of the inner product
between these two states as
\be
|\langle\langle r | l \rangle\rangle| \equiv e^{-\Gamma_{sub}} .
\ee
$\Gamma_{sub}$ is then a decoherence functional in its own right, and
is the natural measure of the decoherence caused by the
electromagnetic field's direct response to a superposition of
sub-leading soft currents.

We evaluate the sub-leading contribution \eqref{eq:sub_gamma} in
Appendix \ref{app:integrals_sub}. The result is that the purely sub-leading
contribution to $\Gamma_{dressed}$ is
\be \label{eq:sub_result}
\Gamma_{sub} \approx {e^2 \over 3 \pi^2}\Omega^2 v^2 \tau^2.
\ee
Because $v \equiv l/\tau$, this expression is equivalent to
\be \label{eq:sub_result_final}
\Gamma_{sub} \approx {e^2 \over 3 \pi^2}\Omega^2 l^2 \equiv {e^2
\over 3 \pi^2} \left({l\over l_{0}} \right)^2,
\ee
where we have used the fundamental length scale $l_{0} \equiv 1 /
\Omega$ defined by our ultraviolet cutoff.

The result \eqref{eq:sub_result_final} is then that the purely
sub-leading contribution to $\Gamma_{dressed}$ has a remarkable form:
it only depends upon the strength of the coupling $e$ of the particle
to the electromagnetic field, and on the size of the interferometer
$l$, measured in units of $l_{0}$, the size of a ``pixel'' in the
implicit coarse-graining of space-time implied by our use of an
ultraviolet cutoff. Unlike the expression \eqref{eq:dressed_result}
for the total amount of decoherence, this purely sub-leading soft
contribution does not depend directly upon the particle's velocity
$v$ or on the time $\tau$ taken to carry out the experiment.

\subsection{Sub-Leading Dressing} \label{sec:subleading_dressing}

Unlike the leading order contribution, the sub-leading contribution
to decoherence is infrared-finite. The sub-leading part however does
arise via the same physical mechanism as the leading part -- from a
boundary term in the electric current. Should we then apply infrared
dressing at the sub-leading order \cite{choi}, just as we did at
leading order?

There are no obviously unphysical divergences which arise in our
calculation at sub-leading order, so it does not seem that the
consistency of the theory itself demands further dressing. Instead,
we should ask whether the question might be able to be settled
experimentally.

To that end, let us consider a specific form of the potential
sub-leading dressing. We will assume that, if such dressing were to
exist, we may apply it semiclassically by simply setting $j_{sub}^a$
to zero everywhere, exactly analogous to the way in which applied the
leading order dressing. Setting both the leading and sub-leading soft
currents to zero leaves only the hard current $j_{hard}^a$ to
contribute to the decoherence functional:
\begin{align} \label{eq:hard_gamma}
&\Gamma_{hard} \equiv   {1\over2}  \int \frac{d^3q}{(2\pi)^3
2\omega}\; P_{ab} \, \delta  j_{hard}^a(q) \delta  j_{hard}^b(-q) =
\\
&\hspace{-1em}{e^2 \over (2\pi)^3} \int_0^\Omega {d\omega \over
\omega} \; \left(   2 - 2\cos\omega\tau  + i\omega\tau
e^{i\omega\tau} - i\omega\tau e^{-i\omega\tau} + \omega^2 \tau^2
\right)  \nonumber \\
&\hspace{-1em}\times\oint dS^2(\hat n) \, \omega^2 \, \left[ 2
\frac{\dot{X}_1\cdot \dot{X}_2}{(q \cdot \dot{X}_1)(q \cdot
\dot{X}_2)} - \frac{1}{(q \cdot \dot{X}_1)^2} - \frac{1}{(q\cdot
\dot{X}_2)^2} \right] , \nonumber
\end{align}
where the expression for $\delta j_{hard}^a$ comes from eq.
\eqref{eq:dj_parts}.

We evaluate this expression in Appendix \ref{app:integrals_hard},
with the result
\be
 \label{eq:hard_result}
\Gamma_{hard} \approx {e^2 \over 3 \pi^2} v^2 \left[2\ln{\Omega\tau}
+ \frac{1}{2} \Omega^2 \tau^2\right],
\ee
in the limit $\Omega\tau \gg 1$. Comparing this result to the
prediction \eqref{eq:dressed_result} after only the leading dressing
had been applied, we see that applying the sub-leading dressing
increases the predicted amount of decoherence substantially. The
decoherence now grows quadratically with $\Omega\tau$ rather than
logarithmically, and by eqs. \eqref{eq:v_rl} and
\eqref{eq:LR_inner_product}, this implies that the visibility of any
interference effects will decrease accordingly.

We are now faced wth an interesting choice. On the one
hand one can argue that the prediction \eqref{eq:dressed_result} is
consistent with well-known prior work, and that the addition of sub-leading
dressing seems to increase the predicted decoherence drastically
over that well-known result. Then, a conservative point of view would say
that sub-leading dressing is not necessary.
However one can also argue that, in the absence of strong theoretical
arguments, the question of whether sub-leading dressing exists should
be settled by an appropriate experiment -- here, an interferometric
experiment. The marked difference between the predicted decoherence rates
with and without sub-leading dressing suggests that this such an experiment
would not be unfeasible. We stress that such an
experiment must take great care to respect the boundary conditions we
have imposed upon the particle trajectories in our toy model, if it
is to be sensitive to the infrared effects we have studied here.
The importance of this question is such that we need to go beyond toy models
and discuss what a real experiment might look like -- this we now do.


\section{Experimental Implications}
\label{sec:Expt}


Any realistic experiment that tries to observe
the presence or absence of sub-leading soft dressing will
be fraught with ``real world'' complexities. This is partly because
the purely photonic decoherence depends very much on the the sample
geometry. But even more important is that in any real experiment
there are lots of other decoherence sources. A proper discussion of
these would inevitably occupy many papers.

Accordingly, in what follows we try to satisfy a more limited goal,
viz., (i) estimating the soft photon decoherence effects arise in
two model experiments, and
(ii) estimating the size of the other decoherence effects which
will also appear. To focus the discussion we consider two
specific experiments, viz., a 2-slit diffraction experiment with
electrons, and an interferometric experiment with massive particles.
We do not do any really detailed calculations, but rather indicate
what will be involved in such calculations, and what kind of
answers one expects.

\subsection{2-slit System}
\label{sec:elec-2slit}

\begin{figure}[ht]
\centering
\includegraphics[scale=0.25]{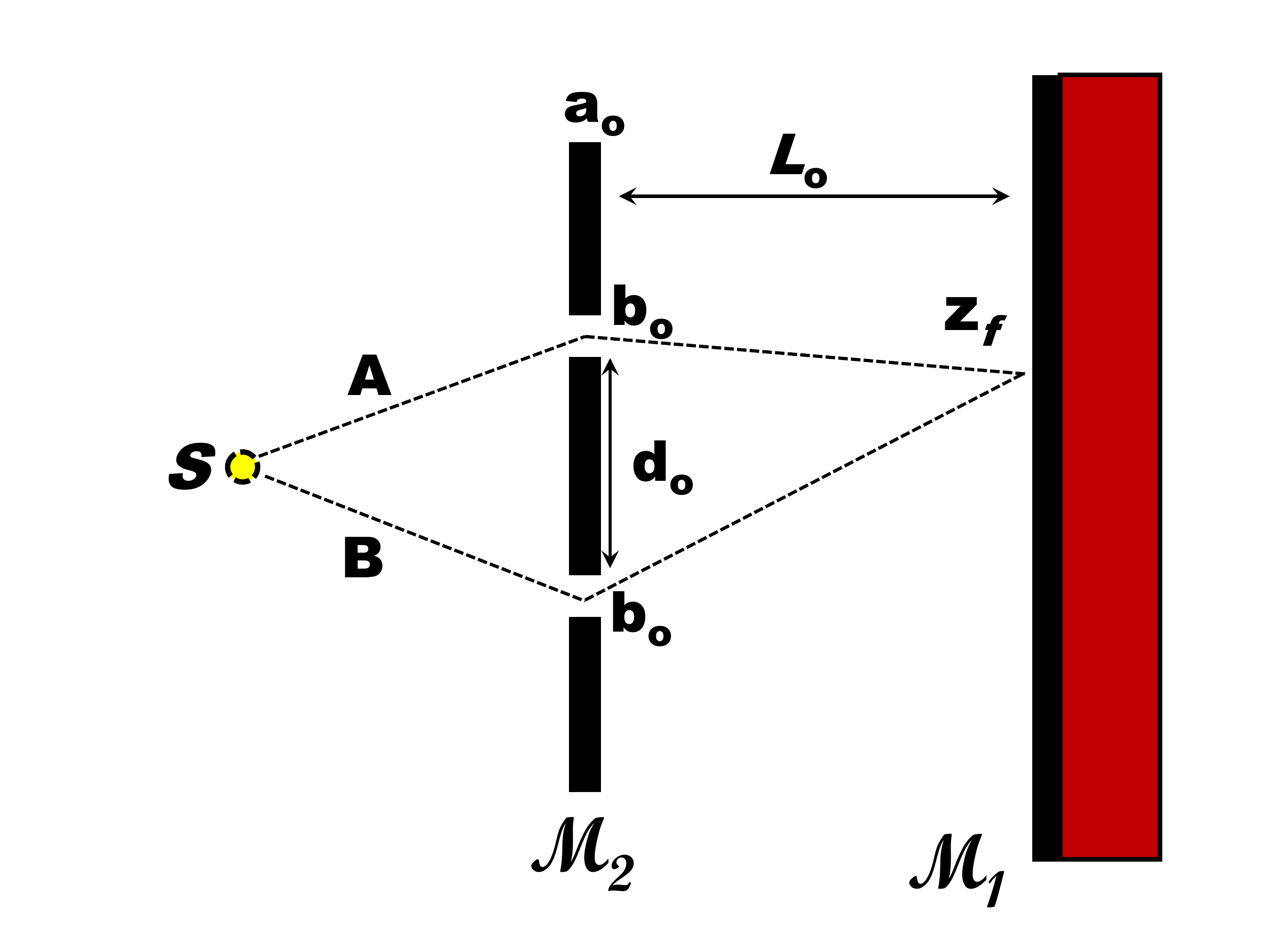}
\vspace{-3mm}
\caption{An idealized view of a 2-slit interference system. An object
${\cal S}$ can follow either path A or path B through the slit system
${\cal M}_2$ to a final coordinate $z_f$ on the screen ${\cal M}_1$.
the plate ${\cal M}_2$ has thickness $a_o$; each slit has width $b_o$,
with distance $d_o$ between slits, and distance $L_o$ from ${\cal M}2$
to ${\cal M}_1$.}
 \label{fig:2slit}
\end{figure}

The 2-slit experiment for electrons is of course well known
\cite{feynman3}; however the real physical processes taking place
during the passage of an object through the slit regions are
actually quite complex. To focus the discussion, consider the
idealized setup shown in Fig. \ref{fig:2slit}.


\begin{figure}[ht]
\centering
\includegraphics[scale=0.25]{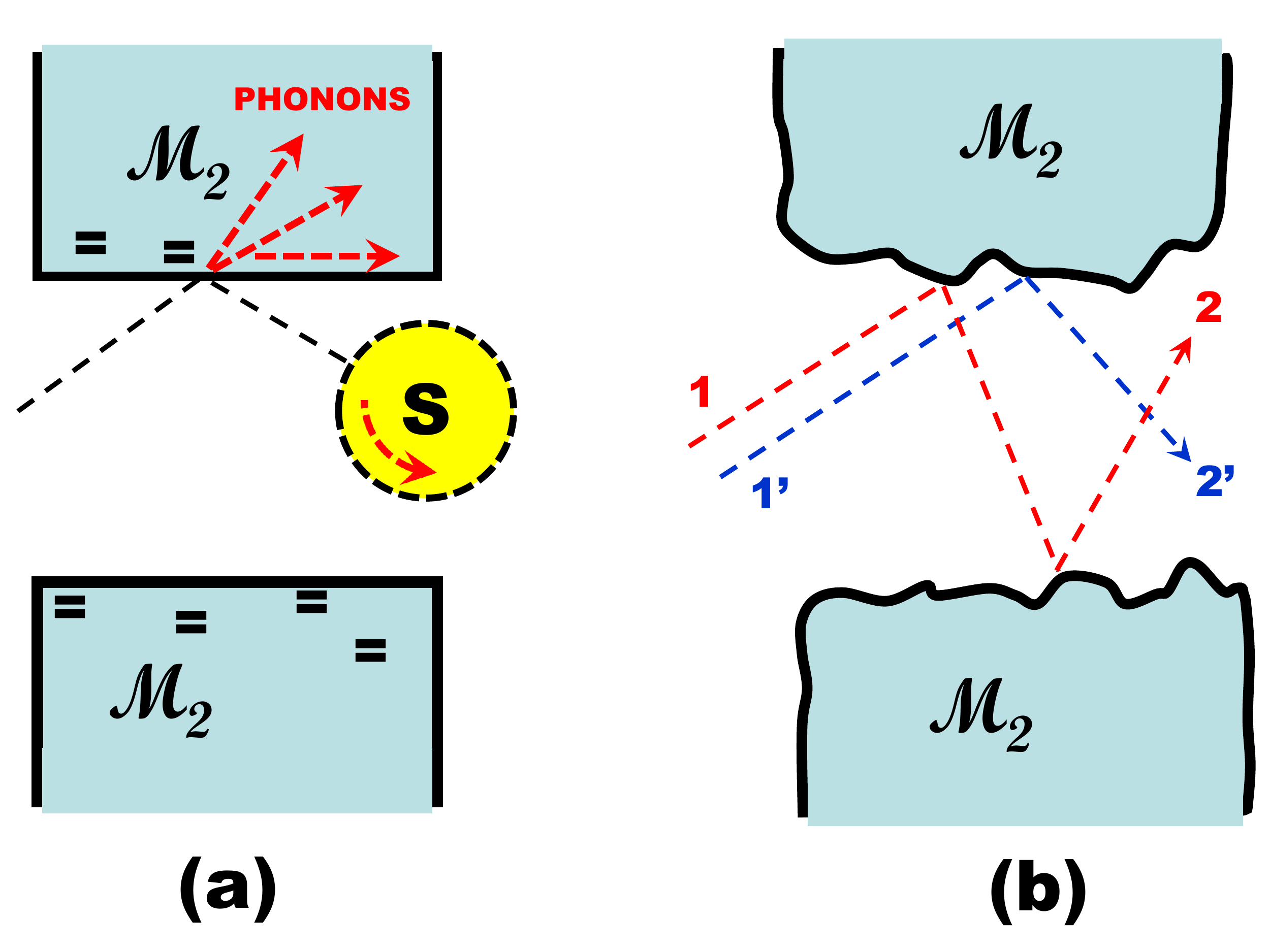}
\vspace{-2mm}
\caption{ In (a) we show some of the scattering processes occuring when
an extended body ${\cal S}$ passes through an idealized slit. Collision
between ${\cal S}$ and the slit can create bulk or surface phonons,
which
can, e.g., set ${\cal S}$ into rotation. It can also interact with
2-level
systems in ${\cal M}_2$. In (b) we show paths $1 \rightarrow 2$ and $1'
\rightarrow 2'$ for
2 incoming systems with identical momentum and energy, but displaced
from each other. We will reject any experimental design in which such
proceses are important. }
 \label{fig:localSlit}
\end{figure}

The object ${\cal S}$ passing through this idealized system may be
an electron, which typically will have a long wavelength $ \lambda$
compared to either the slit width $b_o$ or the thickness $a_o$ of the slit
plate ${\cal M}_2$, ie., $\lambda \gg a_o, b_o$. Alternatively we can
consider some larger object, such a a large molecule, or, e.g., an
insulating nanoparticle made of SiO$_2$, or a conducting metallic
nanoparticle - in this case one may have $\lambda \ll a_o, b_o$.
Experiments are in principle possible in both limits, but the physics is
quite different in the two cases.
In a real 2-slit system of this kind surface irregularities may be
important in the short wavelength limit (compare Fig. \ref{fig:localSlit}).
The irregular shape of the
surface can be modelled in various ways; typically one uses a disordered
scattering potential and averages over the disorder. Note that this is not a
decoherence mechanism -- however it will cause unwanted overlap of the L and R
electron states. We will therefore treat it as an unwanted feature of the
experiment, and assume that a well-designed experiment will have eliminated
such irregularities.
In the long wavelength limit $\lambda \gg a_o, b_o$, each slit effectively
re-radiates the electron
states, and one can analyze this using the usual treatment \cite{ford}.
If we are to find the experimental decoherence rate for this system, we
need to understand the role of mechanisms other than the long wavelength
photon decoherence mechanism we are looking for. These include:

(i) Discrete surface degrees of freedom
associated with unpaired surface electrons (``dangling bonds''), or other
dynamic impurity or defect modes which may be on the surface or in this
bulk. These degrees of freedom are usually modelled as a ``two-level
system'' environment \cite{RPP00,PWA76}; on the macroscopic scale a
large set of such fluctuating 2-level systems tend to show up
experimentally as fluctuating ``patch  potentials''
\cite{patchP}. The effect of these defects is to (a) electrostatically
perturb the paths of the electrons, and (b) cause decoherence in
their dynamics, which can be modelled using a ``spin bath'' model
\cite{RPP00} for the coupling to this environment. At low temperatures this is often
the main source of environmental decoherence in the motion of ${\cal S}$.

(ii) both surface and bulk phonons in the slit system can interact
with ${\cal S}$ as it passes through the slit, with energy and momentum
exchange between ${\cal S}$ and the slit system ${\cal M}_2$.  If the
slit system is conducting, then the electron motion can also excite
gapless surface electronic excitations. The effect of interaction with
these modes is cause more environmental decoherence - these effects can
be modelled using an ``oscillator bath'' model \cite{CalLegg83} for
this environment, this oscillator bath decoherence will dominate at
higher energies or higher $T$ over the spin bath decoherence.

(iii) at finite temperatures, ${\cal S}$ will interact with a
thermal bath of photons while traversing the system. In an evacuated
2-slit system, the photon bath temperature will be determined by the
temperature of the walls of the system, and of the slit and screen
system.

The way in which these different effects can alter the dynamics of
${\cal S}$ is shown schematically in Fig. \ref{fig:localSlit}. If we
ignore surface shape irregularities, then we have the situation shown
schematically in Fig. \ref{fig:localSlit}(a), in which both surface and
bulk phonons can be emitted or absorbed by ${\cal S}$, with a
scattering
matrix element $\Gamma(\epsilon, \epsilon'; {\bf k}, {\bf k'})$ between
energy/momentum states $(\epsilon,{\bf k})$ and $(\epsilon', {\bf k'})$
that needs to be determined from microscopic theory (taking into
account that one may emit single or multiple phonon excitations).

Note that ${\cal S}$, if it is an extended object like a molecule or
particle, will have many internal degrees of freedom as well; these may
be rotational, vibrational, or discrete (as in, e.g., spin
or vibron modes). Thus as ${\cal S}$ scatters off ${\cal M}_2$, any of
these modes can be excited, and this will also be incorporated in a
detailed calculation of the scattering amplitude. Note that the
discrete
modes of ${\cal S}$ will not have any momentum quantum number; but they
can interact directly with discrete modes like two-level systems in
${\cal M}_2$. One can also have processes in which ${\cal S}$ interacts
with a two-level system which then recoils (thereby emitting a phonon).

As already noted, in reality the surface will not have the idealized
flat surfaces shown
- Fig. \ref{fig:localSlit}(b) exaggerates these irregularities somewhat. When
the wavelength $\lambda = \hbar/M_ov$ of the centre of mass coordinate of
${\cal S}$, moving at velocity $v$, satisfies $\lambda \ll a_o, b_o$,
then the scattering will be sensitive to these irregularities, so that
two given objects ${\cal S}$ and ${\cal S'}$ coming in with identical
$(\epsilon, {\bf k})$, but different impact parameters, will be
scattered differently. On the other hand for low-energy electrons
$\lambda \gg a_o, b_o$, so that surface irregularities will be unimportant.

As already noted, we will reject any experimental design for massive objects
(such that $\lambda \ll a_o, b_o$) if surface irregularities do play any role.


The decoherence mechanisms just discussed are
distinct from those that are the prime focus of this paper: but in any
experiment they will also contribute, and their detailed theoretical
treatment will obviously be complicated. To get some idea of how things
may work, we now consider the 2 examples of interest, in turn.

\subsection{Electron interferometry}
\label{sec:slitGeom}

We begin with electrons, which have a very low mass, and hence long
wavelength. In early discussions of the 2-slit system, going back to
Heisenberg, Einstein and Bohr, the actual electron-slit interaction is
not
analyzed. One instead simply assumes total momentum conservation so
that the deviation in the electron path as it goes through one or
other slit is accompanied by a recoil of the slit system. This
analysis however does not allow us to say anything about the
photon-mediated electron
decoherence, which depends in principle on the detailed path of the
electron (and in particular, the acceleration it undergoes).

We can certainly make a crude estimate of the photon-mediated
decoherence rate, by
noting that for the geometry shown in Fig. \ref{fig:2slit}, one
can assume that the electron acceleration, caused by interaction with
the slits, takes place over a length scale $\sim a_o$, the thickness
of the plates in which the slits are situated. Then, if the
slit-screen distance is $L_o$, and the distance between the slits is
$d_o$, the acceleration suffered by the electron as it passes through
one or other of the slits is given by
\begin{eqnarray}
a_j \; \sim \; {\delta v_j \over \delta t} \; &\sim& \; {v^2 \over
\ell_o} \theta_j \nonumber \\
&\sim& \; {v^2 \over \ell_o} (z_f \pm \tfrac{1}{2} d_o)
 \label{a-j}
\end{eqnarray}
where $j = A,B$ labels the two paths for the electron, $z_f$ is the
final position (on the $\hat{z}$ axis) of the electron on the screen,
$\theta_j$ is the angle through which the electron is deflected on
the $j$-th path, and $v$ is the electron velocity as before.

In this case the the frequency $\Omega = 1/\Delta t$, where $\Delta t
= a_o/v$ is just the time taken for the electron to pass through the
interaction region; and the ``coarse-graining'' lengthscale $l_0$
defined above can also be taken to be $a_o$ (in conventional units).

Let us consider a charged particle with charge $Qe$, where $e$ is the
electronic charge. Restoring the units to MKS units, so that the fine
structure constant $\alpha =\frac{e^2}{4\pi}$, we then have, for this experiment,
the following predictions from eqs.
(\ref{eq:dressed_result}) and (\ref{eq:hard_result}) for
the decoherence rates:
\begin{equation}
\Gamma_{dressed} \;\sim\; Q^2 {16\alpha \over 3\pi}
\left(\frac{v}{c}\right)^2 \; \ln(L_o/a_o)
\end{equation}
\begin{equation}
\Gamma_{hard} \; \sim \; Q^2 {8\alpha \over 3\pi}
\left[2\ln(L_o/a_o) + \frac{1}{2}(L_o/a_o)^2 \right]
\end{equation}

\vspace{3mm}

Are these estimates changed by a more sophisticated analysis?
It turns out that this is not a simple question: the mechanism by
which momentum and energy is exchanged between the electron, the
slit system, and photons is still being debated \cite{batelaan16}.
Here we give our point of view.

In a real 2-slit system, the electron polarizes the 2-slit system as
it approaches it; this polarization can be understood, and the
electron-slit interaction can be written in terms of the dielectric
function of the slit system \cite{echenique81}. However, electrons
that succeed in propagating through the slits will not typically make
contact with ${\cal M}_2$ (if they do, then their paths will be
severely
disturbed, and they are likely to stick to/be absorbed on the surface
of ${\cal M}_2$).

In reality the electrons will interact indirectly with the slit
system through their interaction with the electric fields which exist
in the vicinity of the slits \cite{batelaan16,wessel-berg01}. To
treat this problem properly requires detailed treatment of both the
static and fluctuating electromagnetic fields in and around the slits
-- it is actually an example of the ``Casimir'' problem for this
geometry \cite{casimir}.

It is well known that a proper treatment of even
a simple geometry for such Casimir problems, for a general dielectric
material making up the slits, is extremely complex. A simplified
treatment
\cite{wessel-berg01} starts from the classical ``open cavity''
electromagnetic modes in the slit region, comprising both localized
modes confined to within the slits, and evanescent modes extending away
from the slits.

\begin{figure}[ht]
\centering
\includegraphics[scale=0.25]{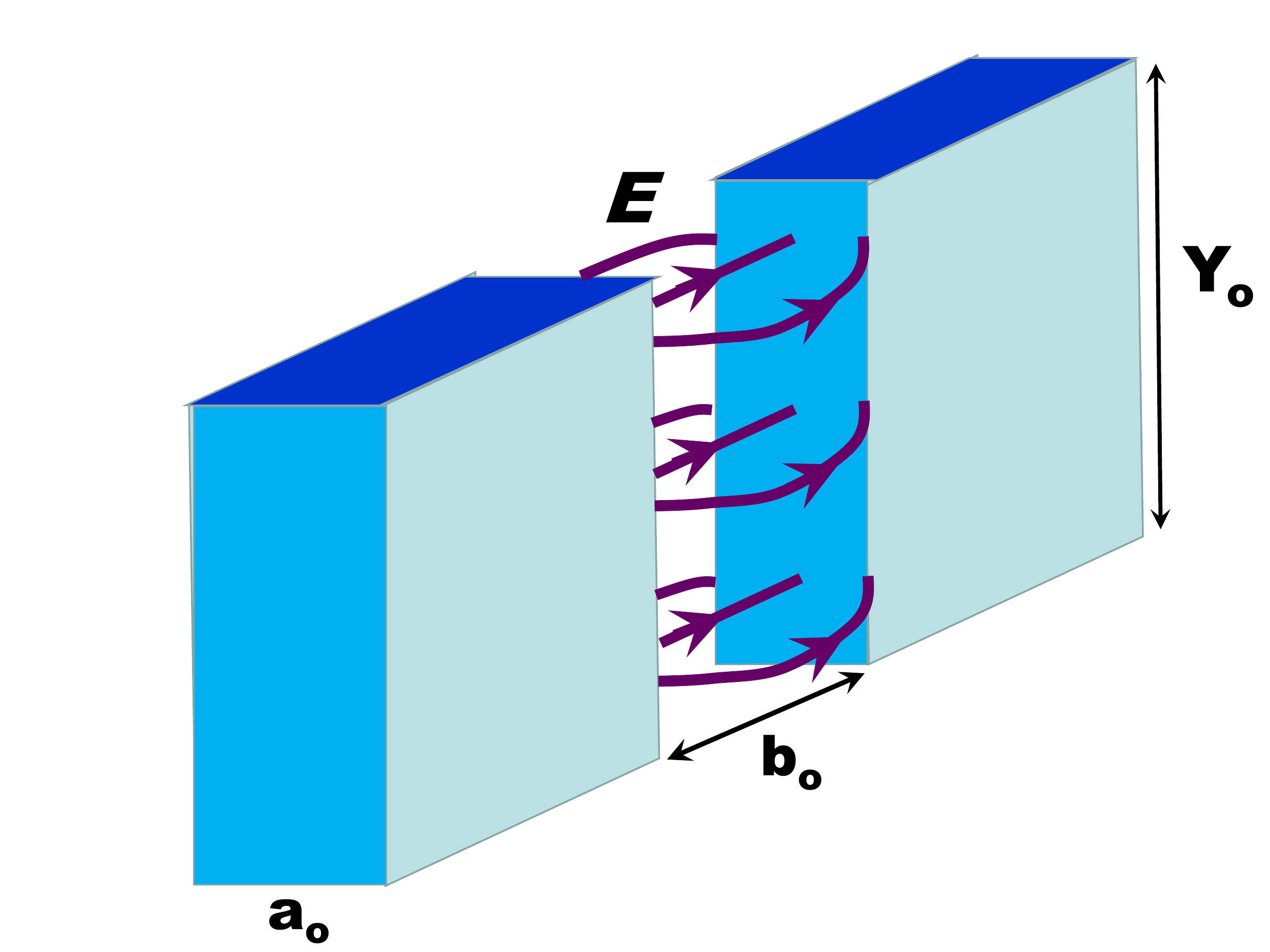}
\vspace{-1mm}
\caption{ Depiction of the electric field configuration around a slit
in
a dielctric slit system ${\cal M}_2$. In a quantum teatment, local
photon
modes exist in this region, and an incoing electron can scatter such
bound-state modes between different energy levels. }
 \label{fig:slitEM}
\end{figure}

To see how this works, consider the geometry shown in Fig.
\ref{fig:slitEM},
in which the slits have width $b_o$, length $Y_o$ and thickness $a_o$.
Then the localized modes, confined to the slit region, have electric
fields of form \cite{wessel-berg01}
\begin{equation}
E_z(x,y) \;=\; E_o \sin \left({\pi y \over Y_o} \right) {\cosh
\kappa_{\omega} x  \over \cosh (\tfrac{1}{2} \kappa_{\omega} a_o)}
 \label{Ez-xy}
\end{equation}
where the wave-vector $\kappa_{\omega}$ is given by
\begin{equation}
\kappa^2_{\omega} \;=\; \left( {\pi \over Y_o} \right)^2 - {\omega^2
\over c^2}
 \label{kappa}
\end{equation}

We then end up with a set of photon bound states in the slit region,
which, when quantized, are populated thermally by photons. Quantized
momentum exchange between the electron and the slit photon modes then
leads to
deviation of the electron path as it travels through the slit,
accompanied by multiple transitions of photons between the bound
state levels. The details are rather lengthy in the general case
\cite{wessel-berg01}, but the key conclusion is that in situations
where photons in many different levels are arrived, one recovers the
results given above for the photon-mediated decoherence rate.


Finally, let us note that in any real experiment the photon bath will
be at some finite temperature $T$. One can in fact generalize all of
the calculations in the previous sections of this paper to finite
$T$. Since the basic methods for doing this are well known (see, e.g.,
refs. \cite{popov,bellac} for general discussions), we simply give
the main results here:

(i) the easiest way to set up this kind of calculation is to define a
decoherence functional \cite{decoF,feynmanV63} for the effect of a
finite-$T$ photon bath on the electron dynamics
\cite{breuer,jordan18}. This then modifies eq. \eqref{eq:gamma_m}
to
\be
\label{eq:gamma_T}
\Gamma[j_L, j_R] =   {1\over2}\int \frac{d^3q}{(2\pi)^3 2 \omega} \;
P_{ab} \; \coth (\beta \omega/2) \; \delta  j^a(q) \delta  j^b(-q),
\ee
where $\beta = 1/kT$, and we now have to assume a specific reference
frame in which the photon bath, at equilibrium temperature $T$, is at
rest - this will usually be determined by the 2-slit system itself,
which is assumed to be in equilibrium with the photon bath. We note
that the same $\coth (\beta \omega/2)$ factor will enter into the
integrals in, e.g.,  eqs. (\ref{eq:softDeco}), (\ref
{eq:gamma_dressed}),    (\ref{eq:gamma_eval_1}),
(\ref{eq:sub_gamma}), and
(\ref{eq:hard_gamma}).

The essential physical effect of the temperature factor here is to
introduce a new energy scale in the problem. If $kT \ll \Omega$ this
becomes a new IR energy scale; in the opposite case where $kT \gg
\Omega$ the electrons are simply interacting with a high-temperature
photon bath.

Finally, one can ask about other decoherence sources in this
experiment.
In fact these will be rather small for a slow-moving electron - the
available phase space for phonon creation will be extremely small at
energies equal to the electron kinetic energy. For an electron
moving at $v = 10^3$ m/sec, having wavelength $\lambda \sim 10^{-7}$m,
this energy is $\sim 40$mK, and correspondingly much less for lower
velocities. There will also be possible decoherence from electron spin
flip processes, mediated by paramagnetic impurities in the slit system,
which for well-prepared slit systems can be neglected.

We thus conclude that the results found in eqs. \eqref{eq:dressed_result} and
\eqref{eq:hard_result}, for photon-mediated
decoherence, are in principle measurable.

\subsection{Electron Interferometry with massive particles}
\label{sec:interferM}

There are well-known experiments \cite{arndt} in which rather massive
molecules are used instead of electrons in 2-slit experiments.
However any attempt to analyze these theoretically is very
complicated \cite{scatt}, because the molecules have irregular
shape, and in general carry angular momentum. Both the angular
momentum and the internal ``shape'' degrees of freedom of the molecule
(modelled by vibrational and ``twisting'' modes, amongst others) can
couple to the slit system \cite{stamp}, and this
leads to extra sources of ``3rd-party decoherence'' (since the coupling
to these degrees of freedom depends on which path the object takes).

The key differences between such experiments and the electron
interference
experiments described above are (i) the much larger mass and energy,
and much shorter wavelength, associated with the centre of mass; (ii)
the much larger role of other kinds of environmental decoherence; and
(iii) the possibility of using neutral (dielectric) particles; even a
neutral object still interacts with photons if its dielectric
properties are different from the vacuum.

In the following we will only consider electrically neutral particles -
if the particle is charged, we can adapt the previous results for the
electron to estimate the photon decoherence.

In all cases, what we wish to know is -- how does the photon decoherence
compare with other environmental sources of decoherence? There are many
different possible situations -- here we summarize some key ideas.

\subsubsection{Basic Design}
\label{sec:design}

\begin{figure}[ht]
\centering
\includegraphics[scale=0.25]{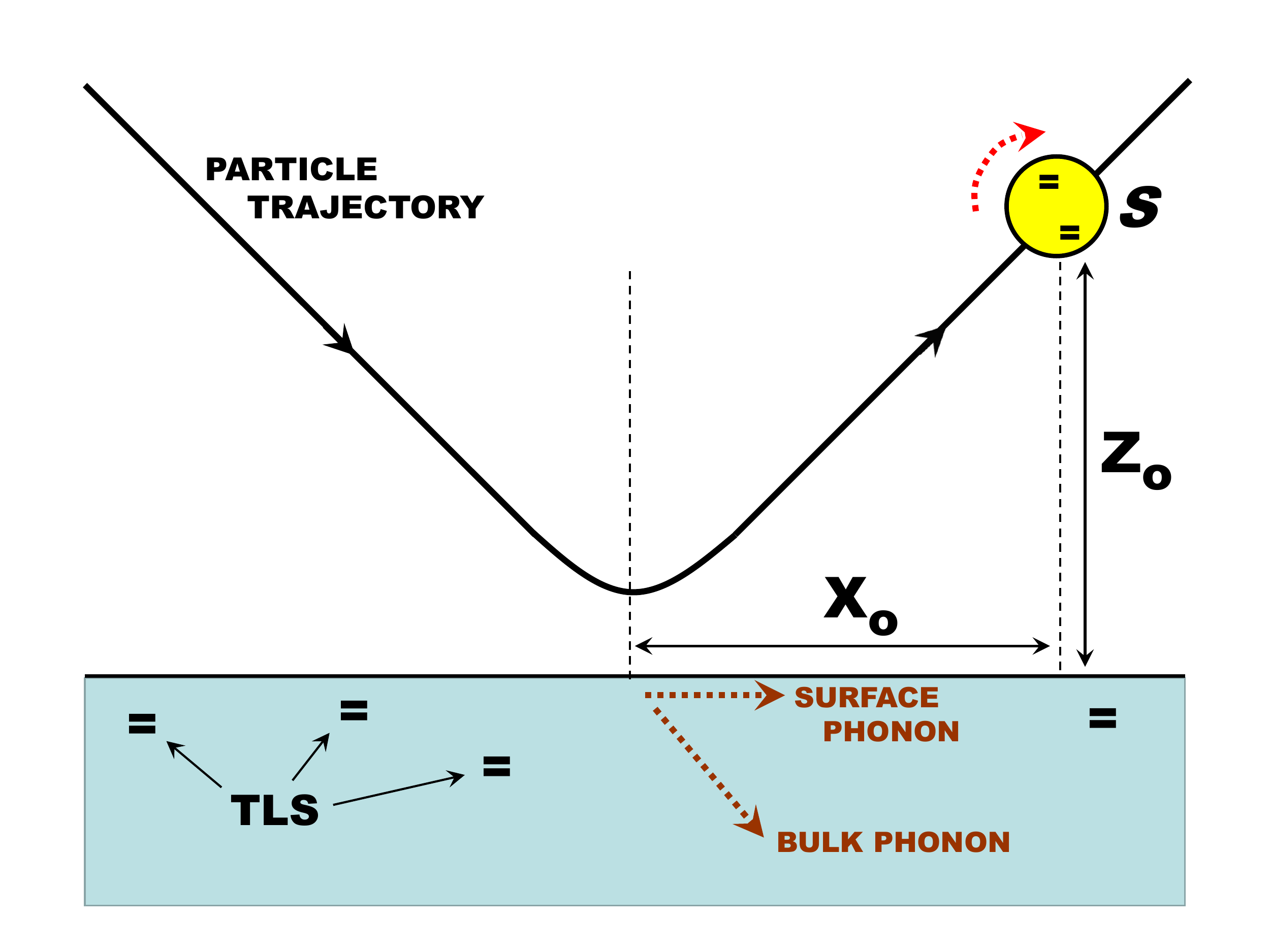}
\vspace{-2mm}
\caption{ Interaction of a massive spherical particle, of
radius $r_o$, with a flat mirror. The coordinates of the centre
of mass of the particle relative to the origin are
$(X_o, Y_o, Z_o + r_o)$; the particle is moving in the
$\hat{x}$-direction. The interaction will generate
phonons in the mirror and vibrational and rotational motion
in the particle. Two-level systems and spins in the particle
can interact with those in the mirror via dipolar interactions. }
 \label{fig:vdW+mirror}
\end{figure}

We rely in what follows on the experimental possibility of almost
elastic reflection of a slow moving particle from a flat mirror. If
both the mirror and the particle are electrically neutral, then the
interaction between them is well known to be a combination of a
short-range repulsion between atoms and a long range van der Waals
attraction. This experiment is shown schematically in Fig.
\ref{fig:vdW+mirror}. If the particle is charged then the charge will
also induce a mirror charge on the mirror. If the particle and mirror
are conducting we have a quite different situation again. Such problems
have been studied for many years
\cite{klimch09,reynaud08,kardar09,reynaud12}.

The most conceptually simple case is of perfectly conducting (but
uncharged) particle and mirror; then the following results are known to
be the case:

(i) When the particle is far from the plane, so that $Z_o \gg r_o$,
one has the asymptotic relativistic van der Waals behaviour
\begin{equation}
V_o(Z_o) \; \sim \; U(Z_o) - {9 \hbar c \over 16\pi} {r_o^3 \over
(r_o + Z_o)^4}     \qquad (Z_o \gg r_o)
 \label{Cas1}
\end{equation}
where $U(Z_o)$ is a short range interaction, which in this simple model
takes the form $U(Z_o) = U_o \theta(-Z_o)$. ;

(ii) on the other hand when $Z_o \ll r_o$ one finds
\cite{vdWaals}
\begin{equation}
V_o(Z_o) \; \sim \; U(Z_o) + \left( {1 \over 3} - {5 \over \pi^2}
\right) {\hbar \pi^3 c \over 720} {1 \over Z_o}
 \label{V-vdWaals}
\end{equation}

In the intermediate range one has $(V_o(Z_o) - U(Z_o)) \sim
O(1/Z_o^3)$ (the long-range non-relativistic van der Waals
behaviour).

In this simple example one already sees the key difference between the
two parts of
$V_o(Z_o)$. The long-range van der Waals term can be determined
via macroscopic considerations (although it can depend in a very
complicated way on the shape and material properties of the two
bodies). This physics has been studied in considerable detail
\cite{vdWaals}.

The result for the short-range $U(Z_o)$, on the other hand, is just the
standard Casimir result for perfect conductors. However, it is
extremely unrealistic for any real conducting system. In reality, a
close approach between 2 large systems like this brings into play
physics at atomic length scales, and energies as much as hundreds of
$eV$.  When a massive particle approaches a surface with appreciable
kinetic energy, a serious distortion of the surface regions of both
particle and surface ensues during the collision, involving the surface
and also layers well below the surface, so that $U_o(Z_o)$ operates
indirectly over length scales ranging from atomic scales up to several
nanometers.  Simulations of even quite simple examples \cite{cleary10}
show that this process is highly complex. In real systems the collision
can
also create defects and dislocations, or move existing ones around (and
as noted, on macroscopic length scales this is seen as a patch
potential \cite{patchP}. It can also involve severe disturbances of the
outer shells of atoms, including ionization, with the liberation of
photons.

Thus for any real system, the physics for a large particle approaching
a plane is extraordinarily complex. The simple model of a perfectly
conducting system, while being a nice simple model to analyze, is far
from being realistic.

We can instead make progress here by considering the interaction
between a solid spherical dielectric and a solid semi-infinite
dielectric system restricted to a half-plane. We assume again that the
closest distance between the sphere, of radius $r_o$, and the solid
half-plane, is $Z_o$.


We begin with an action of form
\begin{equation}  \label{surfAct}
S[{\bf R}_o, h_{\bf q} ] \;=\; S_o [{\bf R}_o] + S_{surf} [h_{\bf q}] +
S_{int} [{\bf R}_o, h_{\bf q}]
\end{equation}
where $h_{\bf q})$ is the Fourier transform of the height fluctuations
of the mirror surface, ie., where $h({\bf r}_{\perp}, t) = h_{\bf q}
e^{i({\bf q}\cdot {\bf r}_{\perp} - \omega t)}$, with a displacement
${\bf r}_{\perp}$ on the surface relative to the origin. The particle
centre of mass coordinate is ${\bf R}_o = (Z_o + r_o, X_o)$  relative
to the same origin on the surface. The various terms in $S[{\bf R}_o,
h_{\bf q} ]$ are then
\begin{eqnarray}
S_o [{\bf R}_o] \;=\; \frac{1}{2}\int dt [ M_o \dot{R}_o^2 - V(Z_o)]
\nonumber \\
S_{surf} [h_{\bf q}] \;=\; \frac{1}{2}\int dt [{\rho_o \over q}
(\dot{h}_{\bf q}^2 - \omega_{\bf q}^2 h_{\bf q}^2)] \nonumber \\
S_{int} [{\bf R}_o, h_{\bf q}] \;=\; - \int dt \sum_{\bf q}
\lambda_{\bf q} ({\bf R}_o) h_{\bf q}
  \label{surfAct2}
\end{eqnarray}
where $\rho_o$ is the density of the mirror, and $\omega_{\bf q}$ is
the dispersion for the surface waves of the mirror -- there will be 2
branches coming from longitudinal and transverse surface phonons. To
find the interaction $\lambda_{\bf q} ({\bf R}_o)$ one assumes
\cite{echenique76} a van der Waals interaction $-g_o/|{\bf r} - {\bf
r'}|^6$ between unit volumes of dielectric situated at ${\bf r}, {\bf
r'}$.

Integrating over the sphere and the mirror
\cite{echenique76,edwards88}, introducing oscillator coordinates
\begin{equation}
x_{\bf q} \;=\; (\hbar/2m_q\omega_{\bf q})^{1/2} h_{\bf q}
 \label{x-q}
\end{equation}
we can finally write the effective action for the particle/mirror
system as $S_{eff} = \int dt L_{eff}$, where $L_{eff}$ has the
Caldeira-Leggett form
\begin{eqnarray}
L_{eff} \;=\; \tfrac{1}{2} M_o \dot{R}_o^2 - V(Z_o) + \nonumber \\
\tfrac{1}{2} \sum_{\bf q} m_q(\dot{x}_{\bf q}^2 - \omega_q^2 x_{\bf
q}^2) - \tilde{\lambda}_{\bf q}({\bf R}_o)  x_{\bf q}
 \label{CL-surf}
\end{eqnarray}
in which the coupling takes the final form
\begin{equation}
\tilde{\lambda}_{\bf q}({\bf R}_o) \;=\; -{2^{1/2} \pi^2 \over 3}r_o^3
g_o {q^2 \over Z_o^2} K_2 (qZ_o) \cos(qX_o)
 \label{tilde-Lam}
\end{equation}

With this result one can calculate phonon decoherence for some path of
the particle \cite{echenique76,edwards88}. To find the photon
decoherence one derives a similar action for the coupling to photons.
Let us now briefly consider what one expects for the different
contributions to the decoherence.

\subsubsection{Contributions to Decoherence}
\label{sec:Cont-deco}

As already noted, if the massive particle is charged, we can estimate
photon decoherence using the results given for the simple electron.
Here we focus on the neutral particle discussed above.

To properly formulate this problem, we note that the
electrodynamic properties of the neutral particle are described by a
dielectric function $\varepsilon({\bf q}, \omega)$. At the energy
scales of interest here, the photon wavelength $\lambda = 1/|{\bf q}|
\gg r_o$, and the effective interaction between photons and the
moving particle will be  proportional to its dielectric
polarizability  $\alpha_p \sim O(\varepsilon_o r_o^3)$. This gives a
(Rayleigh) scattering rate $\Lambda_{\bf q} \propto |{\bf q}|^4
\alpha_p^2$; for a dielectric sphere, having no dielectric moment,
one has \cite{schwinger-EM}
\begin{equation}
W^{EM}_{\bf q} \;=\; {8\pi \over 3} \left( {\varepsilon - 1 \over
\varepsilon + 2} \right)^2 \varepsilon_o r_o^6 |{\bf q}|^4
 \label{dielectric}
\end{equation}
where $\varepsilon$ is the long-wavelength limit of $\varepsilon({\bf
q}, \omega)$.

Both the decoherence from scattering off photons, and that from their
absorption and emission, are then determined by $W^{EM}_{\bf q}$;
insertion of this scattering rate into the decoherence functional
(\ref{eq:gamma_T}) now gives a decoherence rate $\propto r_o^6 d\omega
\omega^5 \coth (\beta \omega/2)$,
which is rather small at low $T$, and
which must compete with the decoherence from both phonons and defects.

The phonon decoherence rate from the coupling to surface phonons can be
calculated directly from the form of the coupling given above in
(\ref{tilde-Lam}); this yields a decoherence rate $\propto r_o^6
d\omega \omega^3 \coth (\beta \omega/2)$, which then dominates over
the photon decoherence rate at low $T$.

Finally, one can discuss the decoherence coming from defects and
paramagnetic impurities in the particle-mirror system. Quite generally
one can discuss this using the spin bath representation of these
objects \cite{RPP00}. Typically they will interact via dipolar
interactions (electric or magnetic), and one can write an influence
functional for their effect on the particle decoherence -- we forego the
details here.

From all this we can conclude several things:

(i) if one is to stand any chance of seeing and measuring the photon
decoherence for such neutral particles, we must be able to separate it
from the phonon and defect contributions, and the phonon contribution
will typically be much larger. To minimize phonon emission, and to rule
out inelastic deformation of the surfaces, the
kinetic energy of the massive particle has to be low, low enough so
that the phonon excitation rate is controllable.
Roughly speaking, we would like this kinetic energy to be $\sim 1$ eV
or less, and the energy exchange between the particle and the surface
needs to be hundreds of times lower.

To get an idea of the numbers, we note that for a SiO$_2$ particle of
mass $M_o = 2 \times 10^4$ AMU, containing 167 SiO$_2$ units, and
having diameter $\sim 2$ nm, a kinetic energy of 1 eV is attained when
$v = 100$ m/sec; for a particle of mass $M_o = 2 \times 10^8$ AMU,
containing $1.67 \times
10^6$ SiO$_2$ units, and having diameter $\sim 40$ nm, a kinetic
energy of 1 eV is attained when $v = 1$ m/sec. Now note that the
length scale over which the particle ``bounces'' will be $\sim O(r_o)$,
thereby involving times $\Delta t \sim r_o/v$ and frequencies
$\Omega \sim v/r_o$; for the 2 examples just given, we have
$\Omega \sim 10^{11}$ Hz for the small particle (ie., $\mu$waves), and
$\Omega \sim 5 \times 10^7$ Hz (ie., RF), for the large particle.

We can therefore conclude by saying that the best kind of experiment to
look at soft photon decoherence is very likely the 2-slit experiment
with electrons, or some similar design. A proper comparison of theory
and experiment will require detailed consideration of the electron-slit
interaction, as well as quantifying contributions from other
decoherence sources.


\section{Discussion} \label{sec:discussion}


The presence of infrared divergences demanded that we apply infrared
dressing at leading order. Our view on the status of possible
\emph{sub-leading} dressing however is still murky. While the leading
and sub-leading currents have much in common classically, quantum
mechanically only the former contributes to obviously unphysical
conclusions. Accordingly, we have speculated here about
whether it might be possible to observe sub-leading soft dressing (or
its absence) using an interferometer.

Interpreting the sub-leading contribution \eqref{eq:sub_result_final}
to decoherence is mostly straightforward.
The full electromagnetic field obtains an amount of
which-path information given by \eqref{eq:dressed_result}. The amount
of this information attributable solely to the difference between the
sub-leading soft radiation emitted by each branch of the matter
superposition is given by \eqref{eq:sub_result_final}, which shows
that the sub-leading soft contribution grows quadratically in $l$,
the size of the interferometer. This makes sense, because sub-leading
soft photons are by definition excitations of the electromagnetic
field with a very long wavelength, and are capable of resolving only
large-scale details of the matter system. The interpretation of
\eqref{eq:sub_result_final} is then: the larger the matter system,
the more capable sub-leading soft photons are of making quantum
measurements of it. For larger experiment geometries, more of the
quantum information broadcast by the charged particle then ends up
being stored in sub-leading soft modes, relative to the hard modes.
Keep in mind of course that we have obtained our results using the
dipole approximation, in which we assume that the wavelength of any
emitted radiation is still much larger than the size of the
experiment. It would be interesting to go beyond this approximation
in future work.

We should also at this point caution the reader that we do not regard
our discussion, here and in in Section \ref{sec:subleading_dressing},
to be complete. The soft currents $j_{div}^a$ and $j_{sub}^a$ have
been shown \cite{me} to encode the leading and sub-leading soft
photon theorems \textit{at tree level}. While the leading soft photon
theorem is already exact at tree level, the sub-leading soft photon
theorem in general receives loop corrections
\cite{loops1,loops2,loops3,loops4}. It then would be prudent to
explore quantum corrections to our simple semiclassical model, if we
hope to fully understand whether any sub-leading dressing is
required, and what form it might take. In the meantime, as in other discussions
of sub-leading soft dressing \cite{choi}, our results beyond leading soft order
may be trusted up to $\mathcal{O}(e^2)$ in a perturbative expansion.

We therefore conclude that the issue of the dressing of sub-leading
contributions requires both more theoretical work, and experimental
testing.

Finally, we note that the framework presented here can likely be
extended to other physical systems. In particular, it would be of
interest to apply these techniques, along the lines of our previous
work \cite{me} to a straightforward extension to the study of
leading, sub-leading, and sub-sub-leading \cite{cachazo}
soft graviton radiation.


\section{Acknowledgements}
\label{sec:discussion}


We thank J. Wilson-Gerow for discussions of this work, which was
supported by NSERC in Canada.

\vspace{40mm}


\begin{appendix}



\section{Dressing, in Detail}  \label{app:dressing}


In this appendix, we will illustrate the usual ``dressed'' formalism
\cite{chung,FK,kibble1,kibble2,kibble3,kibble4} using a scattering
amplitude involving only a single charged particle in QED. This
simple example suffices for our needs, since our interferometry model
too involves only a single charged particle. We then show how the
dressing works in our finite-time model, proving that a minimal form
of infrared dressing can be accomplished simply by setting the
divergent soft current $j_{div}^a$ to zero everywhere.

\subsection{Dressed Scattering}

In textbook QED, the scattering amplitude for a charged particle to
transition from the momentum eigenstate $|p_i\rangle$ to the momentum
eigenstate $|p_f\rangle$, while the photon field transitions from
$|\alpha\rangle$ to $|\beta\rangle$ is
\be
\langle p_f, \beta | \hat{S} | p_i, \alpha \rangle,
\ee
where $\hat{S}$ is the QED S-matrix. Applying infrared dressing to
this formalism amounts to making the replacement
\be
\langle p_f, \beta | \hat{S} | p_i, \alpha \rangle \rightarrow
\langle p_f, \beta | e^{-\hat{R}_{f}} \, \hat{S} \, e^{\hat{R}_{i}} |
p_i, \alpha \rangle,
\ee
where the ``dressing operator,'' which acts on the photon Hilbert
space, is
\be \label{eq:dress_op}
\hat{R}_{i/f} \equiv
 \sum_{h = +, -}  \int \frac{d^3q}{(2\pi)^3 \sqrt{2\omega}} \left[
 F^h_{i/f}(q) \hat{a}^\dagger_h(q)  - \bar{F}^h_{i/f}(q) \hat{a}_h(q)
 \right]   ,
\ee
in which $\hat{a}^\dagger_h(q)$ and $\hat{a}_h(q)$ respectively
create and annihilate a photon with momentum $q^a \equiv
\omega(1,\hat{n})$ and helicity $h$, and obey
\be
\left[ \hat{a}_h(q),   \hat{a}^\dagger_{h'}(q')  \right]  = \delta_{h
h' }(2\pi)^3 \delta^{(3)}(\vec{q} - \vec{q}') .
\ee
We have also defined
\be \label{eq:formfactor}
F^h_{i/f}(q) \equiv e \frac{\epsilon_h(q) \cdot  p_{i/f}}{q \cdot
p_{i/f}} \, \phi(q, p_{i/f}) ,
\ee
with $\epsilon_h^a$ the polarization vector corresponding to the
helicity $h$, and $\phi(q,p)$ may be chosen arbitrarily, except that it
must go smoothly to $1$ as $|\vec{q}| \rightarrow 0$ in order for the
dressing to properly remove the infrared divergence we saw in the
main text.

To remove the divergence, then, it suffices to choose $\phi(q, p)$
nonzero
only in a neighborhood of $|\vec{q}| = 0$, where the divergence
occurs. Such a choice would however necessitate the introduction of at
least one new parameter --
an arbitrary new energy scale ``$\Lambda$'' -- to our model, in order
to
characterize the size of the momentum neighborhood involved in the
dressing. There is no physical energy scale in our interferometry
model with which we could identify $\Lambda$, so instead we make a
more conservative choice involving no new parameters at all, simply
letting
\be \label{eq:min_dress}
\phi(q,p) = 1,
\ee
a choice which has been referred to elsewhere as ``minimal'' dressing
\cite{kapec}. With this choice, the frequency integral in
\eqref{eq:dress_op} should be understood as extending from the
infrared cutoff $\lambda$ up to the ultraviolet cutoff $\Omega$, with
$\lambda$ eventually being taken to zero. We will use this choice of
$\phi(q,p)$ in what follows, and we will see that the minimal
dressing has a very clean interpretation in terms of the infrared
divergent soft matter current $j_{div}^a$, making it a natural choice
in light of the deep interplay between the soft currents and infrared
radiation \cite{me}.

Evolving states using the dressed S-matrix $e^{-\hat{R}_{f}} \,
\hat{S} \, e^{\hat{R}_{i}}$ instead of the bare S-matrix $\hat{S}$
allows one to avoid unphysical infrared divergent decoherence rates.
This has been shown already in scattering calculations
\cite{gordon1}, and we will now show how it works in our finite-time
interferometry model. In particular we want to dress the interaction
picture time evolution operator, which we used in eq.
\eqref{eq:coherent_states}, as
\be
\begin{split}
\mathcal{T} e^{-i \int d^4x \; j^a(x) \hat{A}_a(x)} \\
\rightarrow e^{-\hat{R}_{f}}  \, &\left[\mathcal{T} e^{-i \int d^4x
\; j^a(x) \hat{A}_a(x)}  \right]  \,e^{\hat{R}_{i}} ,
\end{split}
\ee
with $\hat{R}_{i/f}$ given by \eqref{eq:dress_op}, but now with the
factors $F_{i/f}^h(q)$ of \eqref{eq:formfactor} evaluated at the
beginning and end of the interferometry experiment (at proper times
$s_{i/f}$), rather than at the asymptotic times $t\rightarrow \pm
\infty$ relevant in scattering scenarios. We will now show that in
order to effect minimal dressing, using \eqref{eq:min_dress}, we need
only set the divergent part $j_{div}^a$ of the electromagnetic
current $j^a$ appearing here to zero. That is, we will show that
\be  \label{eq:app1}
\begin{split}
e^{-\hat{R}_{f}}  \, &\left[\mathcal{T} e^{-i \int d^4x \; j^a(x)
\hat{A}_a(x)}  \right]  \,e^{\hat{R}_{i}}  \\
 &= \mathcal{T} e^{-i \int d^4x  \;  [j^a - j_{div}^a](x)
 \hat{A}_a(x)} .
\end{split}
\ee
The mechanism by which minimal infrared dressing removes infrared
divergences is thus the wholesale decoupling of the Maxwell field
from the infrared divergent soft current $j_{div}^a$.

\subsection{Minimal Dressing Removes the Divergent Soft Current}

To prove eq. \eqref{eq:app1}, let us begin with the second line and
work our way backward. The second term in the exponent there is
\be \label{eq:app2}
i   \int d^4x \; j_{div}^a(x) \hat{A}_a(x),
\ee
with $j_{div}^a$ defined by eq. \eqref{eq:soft_current_parts}.
Using the mode expansion of $\hat{A}_a(x)$,
\be
\hat{A}_a(x) = \sum_{h = +,-}  \int \frac{d^3q}{(2\pi)^3
\sqrt{2\omega}} [ \epsilon^h_a \hat{a}^\dagger_h(q) e^{iq\cdot x}  +
\bar{\epsilon}^h_a \hat{a}_h(q) e^{-iq\cdot x}   ],
\ee
we can rewrite \eqref{eq:app2} as
\begin{align}
&i \int d^4x \; j_{div}^a(x) \sum_{h = +,-} \int \frac{d^3q}{(2\pi)^3
\sqrt{2\omega}} [ \epsilon^h_a \hat{a}^\dagger_h e^{iq\cdot x}  +
\bar{\epsilon}^h_a \hat{a}_h e^{-iq\cdot x}   ] \nonumber \\
&= i \sum_{h = +,-} \int \frac{d^3q}{(2\pi)^3 \sqrt{2\omega}}  [
\epsilon^h_a \hat{a}^\dagger_h  j_{div}^a(q)  + \bar{\epsilon}^h_a
\hat{a}_h  j_{div}^a(-q)  ]  \\
&= \sum_{h = +,-} \int \frac{d^3q}{(2\pi)^3 \sqrt{2\omega}} [
\bar{\epsilon}^h_a \hat{a}_h   -   \epsilon^h_a \hat{a}^\dagger_h   ]
\frac{\dot{X}^a(s_f)}{q\cdot \dot{X}(s_f) }     - (s_f \rightarrow
s_i) \nonumber .
\end{align}
Comparing to eqs. \eqref{eq:dress_op} and \eqref{eq:formfactor}, this
is nothing but
\be
-\hat{R}_f + \hat{R}_i ,
\ee
with $p^a_{i/f} = m\dot{X}^a(s_{i/f})$ and the choice of minimal
dressing $\phi(q,p) = 1$.

This result allows us to conclude that
\be
\begin{split}
&\mathcal{T} e^{-i  \int d^4x \;  [j^a - j_{div}^a](x) \hat{A}_a(x)}
\\
=\; &\mathcal{T} e^{-i  \int d^4x \;  j^a(x) \hat{A}_a(x) - \hat{R}_f
+ \hat{R}_i}.
\end{split}
\ee
We can rewrite this using the form \eqref{eq:Current} of the position
space current $j^a(x)$ as
\be
\mathcal{T} e^{-i e \int_{s_i}^{s_f} ds \;  \dot{X}^a(s)
\hat{A}_a(X(s)) - \hat{R}_f + \hat{R}_i},
\ee
in order to see explicitly that the time ordering of operators using
the coordinate time $t$ gives the same result as ordering operators
using the proper time $s$.
Because the dressing operators here act at the beginning and end of
the experiment -- i.e. $\hat{R}_{i/f}$ contains a contribution only
when the proper time is $s_{i/f}$ -- we can thus pull $\hat{R}_f$ out
from under the time ordering operator $\mathcal{T}$ to the left, and
we can pull $\hat{R}_i$ out to the right, leaving
\be
e^{-\hat{R}_{f}}  \, \left[ \mathcal{T} e^{-i e \int_{s_i}^{s_f} ds
\;  \dot{X}^a(s) \hat{A}_a(X(s)) } \right]  \,e^{\hat{R}_{i}}    ,
\ee
or equivalently
\be
e^{-\hat{R}_{f}}  \, \left[\mathcal{T} e^{-i \int d^4x \; j^a(x)
\hat{A}_a(x)}  \right]  \,e^{\hat{R}_{i}}  ,
\ee
proving eq. \eqref{eq:app1}.


\section{Integrals}


This appendix includes further detail on the evaluation of the
dressed decoherence functional $\Gamma_{dressed}$, and the
sub-leading soft and hard contributions $\Gamma_{sub}$ and
$\Gamma_{hard}$ to this, which are defined in the main text. Our
approach to these calculations is similar to that of Breuer \&
Petruccione \cite{breuer}.

\subsection{The Dressed Decoherence Functional}
\label{app:integrals_dressed}

First write $\Gamma_{dressed}$ (eq. \eqref{eq:gamma_eval_1}) as
\be \label{eq:dressed_integrals}
\Gamma_{dressed} \equiv {e^2 \over (2\pi)^3} \mathcal{I}_{\omega}
\mathcal{I}_{\hat{n}}
\ee
with
\be \label{eq:Ifreq}
\mathcal{I}_{\omega} \equiv  2\int_0^\Omega {d\omega}  \,
\frac{\left(1 -  \cos{\omega\tau}  \right)}{\omega}
\ee
and
\begin{align}\label{eq:Iangle}
&\mathcal{I}_{\hat{n}} \equiv \\
&\oint dS^2(\hat n) \, \omega^2 \, \left[ 2 \frac{\dot{X}_1\cdot
\dot{X}_2}{(q \cdot \dot{X}_1)(q \cdot \dot{X}_2)} - \frac{1}{(q
\cdot \dot{X}_1)^2} - \frac{1}{(q\cdot \dot{X}_2)^2} \right].
\nonumber
\end{align}
We will evaluate each of these integrals one at a time.

The frequency integral $\mathcal{I}_\omega$ can be written as (using
$\varpi \equiv \omega \tau$)
\be
2\int_0^{\Omega \tau} {d\varpi } \frac{(1 - \cos{\varpi})}{\varpi}  .
\ee
We can then use the following identity \cite{GandR} involving the
cosine integral,
\be
\begin{split}
\mathrm{Ci}(\Omega \tau) &\equiv - \int_{\Omega\tau}^\infty d\varpi
{\cos \varpi \over \varpi} \\
&= \gamma_{EM} + \ln{\Omega\tau} - \int_0^{\Omega\tau} d\varpi {(1 -
\cos\varpi ) \over \varpi} ,
\end{split}
\ee
where $\gamma_{EM} \approx 0.577$ is the Euler-Mascheroni constant.
from which it follows that $\mathcal{I}_\omega$ is
\be
 2\int_0^{\Omega\tau} d\varpi {(1 - \cos\varpi)  \over \varpi} =
 2\left[ \gamma + \ln{\Omega\tau} - \mathrm{Ci}(\Omega\tau) \right].
\ee
In the limit $\Omega\tau \gg 1$ discussed in the main text, the
logarithm dominates this expression, giving the approximate result
\be \label{eq:freq_approx}
\mathcal{I}_\omega \approx   2\ln{\Omega\tau} .
\ee

The angular integral $\mathcal{I}_{\hat{n}}$ is a sum of three terms
with integrals of the form
\be
\begin{split}
&\oint dS^2(\hat n) \, \omega^2 \frac{\dot{X}_A\cdot \dot{X}_B}{(q
\cdot \dot{X}_A)(q \cdot \dot{X}_B)}  \\
= &\oint dS^2(\hat n) \, \frac{(1 - \vec{v}_A \cdot \vec{v}_B)}{(1 -
\hat{n} \cdot \vec{v}_A)(1 - \hat{n} \cdot \vec{v}_B)} ,
\end{split}
\ee
with $A,B \in \{1,2\}$.
The Lorentz-invariance of this integral allows us to evaluate it in
any reference frame. For convenience we can choose the frame in which
$\vec{v}_B = 0$, bringing $\mathcal{I}_{\hat{n}}$ into the simple
form
\be \label{chVI:eq:angle_integral}
\oint dS^2(\hat n) \, \frac{1}{(1 - \hat{n} \cdot \vec{v}_{AB})} =
{4\pi \over  v_{AB}} \tanh^{-1}v_{AB},
\ee
where $v_{AB}$ is the relative velocity,
\be
v_{AB} \equiv \sqrt{1 - {1 \over (\dot{X}_A \cdot \dot{X}_B)^2}}.
\ee
We can use the result \eqref{chVI:eq:angle_integral} to see that the
angular integral \eqref{eq:Iangle} is
\be
\mathcal{I}_{\hat n} = 8 \pi \left[ {1 \over  v_{12}}
\tanh^{-1}v_{12} - 1 \right].
\ee
In the geometry illustrated in Figure \ref{fig:2path}, the relative
velocity $v_{12}$ is approximately
\be \label{eq:relvel}
v_{12} \approx \sqrt{2} v,
\ee
in the case of non-relativistic speeds, $v\ll1$,where again $v \equiv
l / \tau$. We can further approximate
\be
{1 \over v_{12}} \tanh^{-1}{v_{12}} = 1 + {1\over 3}v_{12}^2 + \dots
\approx 1 + {2\over 3}v^2
\ee
In this limit, the angular integral is
\be \label{eq:angle_approx}
\mathcal{I}_{\hat{n}} \approx {16\pi \over 3}v^2 .
\ee

Plugging eq. \eqref{eq:freq_approx} and eq. \eqref{eq:angle_approx}
into eq. \eqref{eq:dressed_integrals} yields eq.
\eqref{eq:dressed_result},
\be
\Gamma_{dressed} \approx {4e^2v^2\over3\pi^2} \ln\Omega\tau .
\ee

\subsection{The Sub-Leading Soft Contribution to the Decoherence
Functional}    \label{app:integrals_sub}

The sub-leading soft part of the dressed decoherence functional,
given by equation \eqref{eq:sub_gamma}, can also be written in the
form of a frequency integral multiplied by an angular integral:
\be \label{eq:sub_integrals}
\Gamma_{sub} \equiv {e^2 \over (2\pi)^3} \mathcal{I}^{sub}_{\omega}
\mathcal{I}_{\hat{n}}
\ee
Here, the angular integral $\mathcal{I}_{\hat{n}}$ is the same one we
have already evaluated, so we can simply use the result
\eqref{eq:angle_approx}. The frequency integral is different,
however; we have
\be \label{eq:Ifreq_sub}
\mathcal{I}^{sub}_{\omega} \equiv \int_0^\Omega {d\omega} \; \omega
\tau^2 = \frac{1}{2} \Omega^2 \tau^2.
\ee

Plugging these results into \eqref{eq:sub_integrals} gives the result
\eqref{eq:sub_result}
\be
\Gamma_{sub} \approx {e^2 \over 3 \pi^2}\Omega^2 v^2 \tau^2.
\ee

\subsection{The Hard Contribution to the Decoherence Functional}
\label{app:integrals_hard}

The hard part of the decoherence functional -- obtained by
``dressing'' away both the leading and sub-leading currents -- is
given by equation \eqref{eq:hard_gamma}. As before, this decoherence
functional may also be written in the form of a frequency integral
multiplied by an angular integral:
\be \label{eq:hard_integrals}
\Gamma_{hard} \equiv {e^2 \over (2\pi)^3} \mathcal{I}^{hard}_{\omega}
\mathcal{I}_{\hat{n}}
\ee
Again, the angular integral $\mathcal{I}_{\hat{n}}$ is the same as
before, and we'll re-use the result \eqref{eq:angle_approx}. The
frequency integral in this case is
\be
\mathcal{I}^{hard}_{\omega} \equiv \int_0^\Omega {d\omega \over
\omega} \; \left(   2 - 2\cos\omega\tau  + i\omega\tau
e^{i\omega\tau} - i\omega\tau e^{-i\omega\tau} + \omega^2 \tau^2
\right).
\ee
Comparing to eqs. \eqref{eq:Ifreq} and \eqref{eq:Ifreq_sub}, we see
that $\mathcal{I}^{hard}_\omega$ can be written as
\be
\mathcal{I}^{dressed}_\omega + \mathcal{I}^{sub}_\omega -
\int_0^\Omega {d\omega} \; \left(  i\tau e^{i\omega\tau} - i\tau
e^{-i\omega\tau}  \right),
\ee
allowing us to re-use more of our previous results. Doing so, and
evaluating the new integral in the third term gives
\be
\mathcal{I}^{hard}_{\omega} = 2\left[ \gamma + \ln{\Omega\tau} -
\mathrm{Ci}(\Omega\tau) \right] + \frac{1}{2} \Omega^2 \tau^2 + 2
\left[ 1 - \cos \Omega\tau \right].
\ee
Only two terms here contribute significantly when $\Omega\tau \gg 1$,
leaving
\be
\mathcal{I}^{hard}_{\omega} \approx 2\ln{\Omega\tau} + \frac{1}{2}
\Omega^2 \tau^2.
\ee

Plugging these results into \eqref{eq:hard_integrals} yields
\be
\Gamma_{hard} \approx {2e^2 \over 3 \pi^2} v^2 \left[2\ln{\Omega\tau}
+ \frac{1}{2} \Omega^2 \tau^2\right],
\ee
in agreement with eq. \eqref{eq:hard_result}.

\end{appendix}





\end{document}